\documentclass[runningheads]{llncs}
\usepackage[T1]{fontenc}
\usepackage{graphicx}
\usepackage{hyperref}
\usepackage{bm}
\usepackage{amsfonts}
\usepackage{subcaption}
\usepackage{xcolor}
\usepackage{booktabs} 

\usepackage[numbers,sort&compress]{natbib}

\newcommand{\bra}[1]{\ensuremath{\langle#1|}}
\newcommand{\ket}[1]{\ensuremath{|#1\rangle}}

\usepackage{geometry} 

\begin{document}
\title{Extreme Learning Machines for Attention-based Multiple Instance Learning in Whole-Slide Image Classification}
\titlerunning{ELMs for Attention-based MIL in Whole-Slide Image Classification}
%
%
\author{Rajiv Krishnakumar\inst{1,2} \and
Julien Baglio\inst{1,2} \and 
Frederik F. Flöther\inst{1,2} \and
Christian Ruiz\inst{3} \and
Stefan Habringer\inst{3} \and 
Nicole H. Romano\inst{3}}
\authorrunning{R. Krishnakumar et al.}
%
\institute{QuantumBasel, Schorenweg 44b, Arlesheim, 4144, Switzerland \and
Center for Quantum Computing and Quantum Coherence (QC2),\newline University of Basel, Petersplatz 1, Basel, 4001, Switzerland
\and
Moonlight AI, Switzerland Innovation Park,\newline Place des Sciences 2, 2822 Courroux, Switzerland
}
\maketitle              
\begin{abstract}
Whole-slide image classification represents a key challenge in computational pathology and medicine. Attention-based multiple instance learning (MIL) has emerged as an effective approach for this problem. However, the effect of attention mechanism architecture on model performance is not well-documented for biomedical imagery. In this work, we compare different methods and implementations of MIL, including deep learning variants. We introduce a new method using higher-dimensional feature spaces for deep MIL. We also develop a novel algorithm for whole-slide image classification where extreme machine learning is combined with attention-based MIL to improve sensitivity and reduce training complexity. We apply our algorithms to the problem of detecting circulating rare cells (CRCs), such as erythroblasts, in peripheral blood. Our results indicate that nonlinearities play a key role in the classification, as removing them leads to a sharp decrease in stability in addition to a decrease in average area under the curve (AUC) of over 4\%. We also demonstrate a considerable increase in robustness of the model with improvements of over 10\% in average AUC when higher-dimensional feature spaces are leveraged. In addition, we show that extreme learning machines can offer clear improvements in terms of training efficiency by reducing the number of trained parameters by a factor of 5 whilst still maintaining the average AUC to within 1.5\% of the deep MIL model. Finally, we discuss options of enriching the classical computing framework with quantum algorithms in the future. This work can thus help pave the way towards more accurate and efficient single-cell diagnostics, one of the building blocks of precision medicine.
\end{abstract}
%
%
\section{Introduction}
\label{sec:intro}
In the last decade, artificial intelligence (AI) and deep neural networks (DNNs) in particular have been applied to a rapidly increasing number of problems across healthcare and medicine, yielding remarkable successes \cite{briganti2020artificial,bohr2020rise}. But as adoption of DNNs in healthcare becomes more mainstream, a structured understanding of these ``black boxes" lags behind the boom in the complexity of their architectures \cite{Jiang2024}. Meanwhile, quantum computing has emerged as a potential companion technology to AI, with applications in biomarker discovery based on data modalities such as electronic health records, omics, and images \cite{flother2024quantum}. The advancement of these extraordinary technologies within healthcare calls for careful investigation of their utility in each new use case and data domain.

Many applications in clinical medicine involve data- and compute-intensive approaches while demanding fast turnaround times. In particular, the field of diagnostics is being transformed by image digitization and biomedical image processing algorithms \cite{lu2021}. Surgical pathology and cytology are fields of image-based diagnostics in which a tissue sample is mounted or smeared on a microscope slide and inspected at high resolution. In this diagnostic approach, digitized whole slide images (WSIs) reveal a wealth of information about cell and tissue morphology, the presence of abnormal cells indicating a malignancy, and the expression of specific biomarkers. Blood smears, in which a sample of peripheral blood is smeared on a glass microscope slide, are common cytological preparations that can be digitized as either a single WSI or a series of single-cell images.

Although WSIs and single-cell images are used by pathologists and hematologists to efficiently make a diagnosis, a systematic annotation of thousands of cells on these images cannot occur due to limited resources. Furthermore, WSIs can be several gigabytes in size, particularly if they originate from cytology samples with different z-layers per slide. The vast number of images, their size, and the scarcity of annotations present computational challenges.

The detection of erythroblasts from peripheral blood is a particularly interesting application of AI to image-based diagnostic workflows. These immature, nucleated red blood cell precursors are normally present in the bone marrow but can migrate to peripheral blood as circulating rare cells (CRCs). The presence of erythroblasts in peripheral blood is generally associated with an increased risk of mortality due to their links to inflammatory processes, bone marrow infiltration by solid cancers, myelofibrosis, hypoxia, and other underlying severe health conditions. On average, the detection of erythroblasts in the peripheral blood can precede death by 1 to 3 weeks \cite{Stachon2002}. In a 12-week study monitoring critically ill patients, the presence of fewer than 200 erythroblasts per microliter of peripheral blood increased the chances of death 4-fold to 43.4\%\footnote{Note that the study\cite{Stachon2007} presents a mortality ratio of 50.7\% for critically ill patients with erythroblasts (34 patients out of 67 in critically-ill condition). It includes patients for which the concentration is also higher than 200 erythroblasts per microliter of peripheral blood. If the mortality ratio is calculated excluding the higher concentration rates, it gives the $(34-11)/(66-14)=43.4\%$ mortality ratio we quote.}\cite{Stachon2007}. Beyond the challenge of detecting a rare cell type, erythroblasts must be distinguished from morphologically similar mononuclear cells like lymphocytes as well as any abnormal blood cells that may be present due to other underlying diseases. Image-based detection of erythroblasts could aid in flagging high-risk patients earlier and more often while leveraging routine diagnostic workflows. But in this application, sensitivity of the algorithm is of utmost importance.

Multiple instance learning (MIL) has emerged as an effective way to extract clinical information from whole-slide images, especially when cell-level annotations are not available \cite{quellec2017multiple,astorino2019multiple}. This is useful when only patient-level annotations exist (e.g. a medical diagnosis), when cell-level annotations are expensive to obtain (e.g. genomic aberrations), or when strict boundaries between classes are difficult to define, as is the case with cell type classifications in blood tissues \cite{Campanella2019,Kimura2019}. Attention-based deep MIL methods have yielded further improvements, allowing models to achieve enhanced sensitivity by capturing the benefit of instance-level interpretability while minimizing error propagation to bag-level predictions \cite{ilse2018attention,shi2020loss,han2020accurate,yao2020whole,lu2021,li2021dual,shao2021transmil,wang2023targeting}. In attention-based architectures, the pooling of the feature vectors in each slide is weighted such that more attention is given to the instances that contain discriminative information. This is particularly relevant when detecting rare cell types or subtle morphological signals: if the pooling were not weighted, one would miss the addition of the few cells which call for diagnostic intervention. Unfortunately, the effect of attention mechanism architecture on the detection limit of MIL models is not well understood. We present a systematic study to quantify the detection limit of a model as we vary common architectural elements of the attention mechanism. 

As an extension of this systematic analysis, we also propose a new architecture where we combine attention-based MIL and extreme learning machines (ELMs). ELMs have been explored with the goal of reducing the number of training parameters, introducing layers in neural networks which are randomly initialized and then frozen before model training. ELMs have the advantage of requiring significantly fewer training parameters, as the training is usually performed only on the last layer of the ELM network, thus drastically decreasing the compute resources required. We investigate a classical attention-based extreme MIL approach and apply our new algorithm on the classification of blood cell mutations using the BloodMNIST dataset~\cite{medmnistv2} as a benchmark. This is also the dataset we use for our systematic analysis of the various attention-based deep MIL models.

Finally, quantum machine learning (QML), a subset of quantum computing, has been a field of intense study over the last years and has also helped elucidate the power and limitations of data and classical ML \cite{schuld2015introduction,biamonte2017quantum,huang2021power}. There is meanwhile strong evidence that at least for some tasks QML is provably better, and there are many research threads that are also looking for heuristic advantages \cite{ramezani2020machine,khan2020machine,krunic2022quantum,liu2024towards}. Therefore, we also sketch the path towards a classical-quantum approach where the classical ELMs are replaced by their quantum counterpart in order to improve the accuracy of the results and reduce further the number of trained and untrained parameters.

The paper is organized as follows. In \autoref{sec:end_to_end}, we present the different steps in the process of multiple instance learning on image classification using erythroblast detection as an exemplary application. The creation of bags of individual images, the image segmentation process, feature extraction, and classification of the bags are described. In \autoref{sec:mil}, we introduce MIL models, starting with the intuitive baseline before introducing attention-based MIL and then our novel algorithm combining ELMs with attention-based MIL. In \autoref{sec:datapreprocessing}, we discuss the data pre-processing, from the raw BloodMNIST dataset to the feature extraction and the creation of the bags. The results of our benchmarks are provided in \autoref{sec:results} before concluding with \autoref{sec:discussion} where we also sketch the steps towards a classical-quantum approach for the ELM part of our novel algorithm.

%
%

\section{End-to-End Process of Multiple Instance Learning on Images}
\label{sec:end_to_end}
The goal of a MIL pipeline is to take in a bag of images and output an overall score corresponding to the bag as a whole. We take the following use case as a running example: we would like to build a MIL model to detect whether or not an individual has at least one erythroblast present in a sample of their peripheral blood. Here we define each blood smear image as a ``bag'' of blood cell images. This reflects the typical output of a hematology image analyzer, which digitizes the blood smear at high resolution, detects individual cells, and saves a series of cropped single-cell images. We focus on erythroblasts because their appearance in peripheral blood is rare and associated with adverse survival in a variety of medical conditions. For erythroblast detection, the gold standard to date is still manual microscopic evaluation, and the performance of existing automated analyzers is still limited by the lower number of cells analyzed compared to WSI and detection performance \cite{Cornet2008}.

\begin{figure}[h]%
\centering
\includegraphics[width=\linewidth]{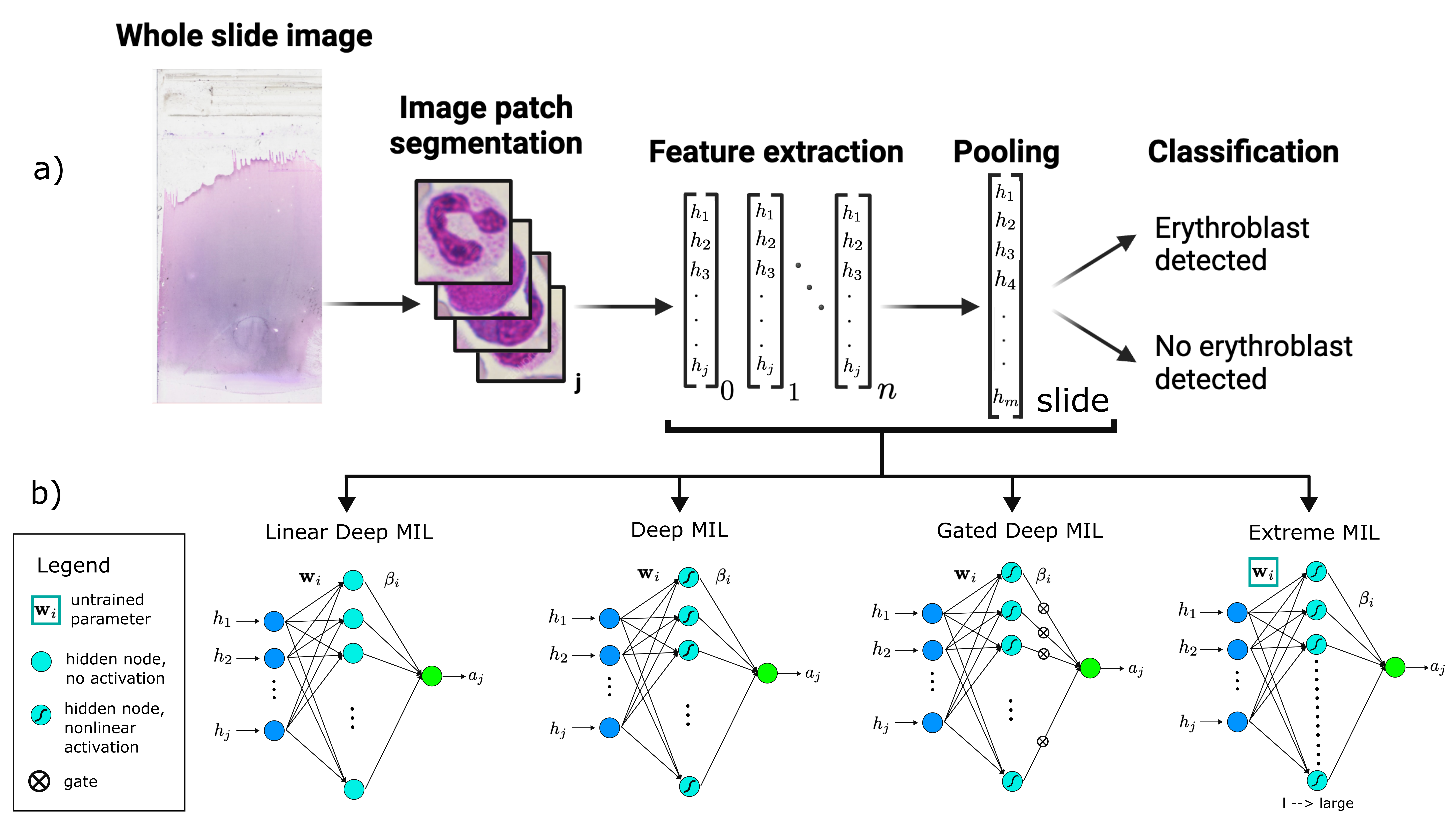}
\caption{a) The pipeline of going from whole slide image to classification and b) a comparison of the attention mechanism architectures investigated. When detecting rare cell types, the performance of a multiple instance learning (MIL) model relies heavily on the pooling operation. Ideally the signal from even a single cell of interest is propagated, regardless of the bag size.}
\label{fig:elm_pipeline}
\end{figure}

There are several steps between digitizing a cytology slide and predicting a classification label. The first step is to create the bag of images that comprises individual object images (\autoref{subsec:bags}). We then convert each individual object image within the bags into a feature vector (\autoref{subsec:feature}). Finally, we pool the feature vectors such that each bag is represented by a single aggregated feature vector, which is then fed to a classifier for inference. As described in \autoref{subsec:aggregation}, the pooling and classification steps can be conceptually thought of as modular stages but are often combined into one step to allow the model to learn more efficiently. An overview of such a computational pipeline is given in the top part of \autoref{fig:elm_pipeline}. We now go through each of these steps in more detail.


\subsection{Bags of Individual Images as Input Data}
\label{subsec:bags}
The term ``bag of images" simply refers to a collection of image patches in which each patch contributes independently to the information in the sample as a whole. In the case of hematology and cytology, the dispersion of cells in a liquid environment makes it possible to use single-cell images as image patches. But in solid tissue pathology where cells are embedded in an environment and their spatial arrangement provides important information, each image patch typically comprises a small field of view containing 20–50 cells on average. Regardless, image patches comprising a bag are explicitly unordered and the number of image patches across bags does not have to be fixed. 

When compiling image patches into a bag, care must be taken to ensure that the contents of the bag represent the minimum information necessary to predict the target label. In the case of rare cell type detection for example, a sufficiently large bag must be compiled to ensure that at least one cell of interest is present in the case of a positive label. On the other hand, the maximum bag size is bounded by the memory capacity of the machine and the type of MIL algorithm implementation.

\subsection{Image Segmentation}
\label{subsec:segment}
When handling gigapixel images, the first step is generally to identify one or more regions of interest. As a result, many pipelines will begin by analyzing a low-resolution whole slide image to discard empty or low-quality regions  \cite{lu2021,hehr2023segmentation}. Once a region of interest is identified, the high-resolution image tiles can be retrieved and further segmented. U-Net is a popular convolutional neural network designed specifically for use with biomedical images, with variations leveraging one-shot detection or vision transformers in order to implement segmentation more efficiently \cite{ronneberger2015segmentation, cao2022unetvit}.

In certain cases we are able to directly obtain a dataset already composed of image patches. In hematology for example, imaging hardware providers such as CellaVision or Vision Hema (West Medica) have made this part of the routine workflow, automatically segmenting individual cell images and delivering the folder of single-cell images as its output. In solid tissue pathology, often the tissue architecture and relative arrangements of cells are informative to the prediction. In this case, one can take advantage of the two-dimensional grid of image tiles that comprise a WSI, using these tiles directly as instances.

\subsection{Instance-Level Feature Extraction}
\label{subsec:feature}
Once the bags have been composed, the next step is to transform each image patch into a feature vector. In the case of embedding-based MIL, this is done explicitly by a dedicated feature extraction network. In instance-based MIL, image patches are transformed directly into class predictions, leaving the task of feature extraction to be performed implicitly within the network.

Feature extraction networks are often subsets of fully-supervised classifiers that have been trained on a relevant proxy task. For example, a network trained to classify cell types might be repurposed for feature extraction by removing its final layers. Recently, self-supervised learning has become a popular method of training feature extractors by training a single model to perform a set of proxy tasks that do not require human-in-the-loop annotation.

\subsection{Classification of Bags}
\label{subsec:aggregation}
In order to arrive at a bag-level classification, a choice must be made regarding how to pool instance-level information. For embedding-based MIL, the feature vectors are pooled into a bag-level representation before being passed to the final layers of the classifier. However, in instance-based MIL pooling of instance-level predictions is performed after the last layer of the model. The former approach, having information-rich feature vectors at its disposal, supports the use of powerful and flexible pooling algorithms such as attention mechanisms. 

At this stage, we pool the feature vectors of every instance into an aggregated bag-level feature vector for classification. It is important to note that we want the bag-level feature vector to be as invariant as possible with respect to the order of the images in the bag. In other words, our classifier should ideally always output the same result for a given blood sample, regardless of the order of the images within the bag. In addition, we want our feature vector to be as invariant as possible with respect to the number of images in the bag. This allows us to vary the bag size as needed (e.g. to take a larger blood sample) while maintaining a positive label if even a single erythroblast is detected. We then use these bag-level feature vectors along with the bag-level labels of the training data to train a classifier. In the next section, we provide more details of existing methods of aggregation and classification as well as discuss our method.

%
%

\section{Multiple Instance Learning Models}
\label{sec:mil}
There are a wide variety of methods to perform the aggregation and classification \cite{bilal2023aggregation}. As mentioned in the previous section, the main criteria for such methods are a) they should not depend on the order of the elements in the bag (i.e. they should be \textit{permutation-invariant} or \textit{symmetric} functions with respect to the elements) and b) they should be as insensitive as possible to the total number of elements in the bag (while still being as sensitive as possible to whether there is at least one positive element amongst the elements in the bag).

First, we consider a very basic method of averaging the features and implementing a logistic regression, presented in \autoref{subsec:logistic}. In addition to describing this baseline model that we compare our deep learning results to, we also take advantage of this section to introduce the notation we are using throughout this paper. We then discuss a more state-of-the art method known as attention-based deep MIL in \autoref{subsec:abdmil} along the modifications that we make to this model. Finally, in  \autoref{sec:abemil} we describe our proposed new model: attention-based extreme multiple instance learning.

\subsection{Simple Averaging and Logistic Regression}
\label{subsec:logistic}
One of the simplest ways to perform the aggregation per bag is to take an average of all the feature vectors. This ensures the invariance with respect to the order of the images and is simple to perform. After this aggregation, the bag-level feature vectors and their corresponding labels can be used to create a logistic regression classifier. We use this method of aggregation and classification as one of our benchmarks and will refer to it as simply the \textit{logistic regression} method from now.

Let us define this method with some rigorous notation, which we can carry forward through the rest of the paper. We define our data as a set of bags, where the $i$th bag contains a set of $n$ feature vectors labeled $\bm{h}^{(i)}_1, \bm{h}^{(i)}_2,...,\bm{h}^{(i)}_n$, each vector represents an image of an object (such as a blood cell), and each bag has a corresponding binary label $l_i$ corresponding to whether or not the bag contains at least one image in the specified category (for instance, the cell type). Here we assume that there is a method to transform each image into a useful feature vector, and we also assume that each bag has the same number of feature vectors without loss of generality. We now define a bag-level feature vector $\bm{z}_i$ as the average of its feature vectors given by

\begin{equation}
\label{eq:aggregation}
\bm{z}_i = \sum_{k=1}^n a^{(i)}_k\bm{h}^{(i)}_k \, ,
\end{equation}

\noindent where $a^{(i)}_k=1/n$ for all $k$. Finally, we use our set of bag-level features $\bm{z}_1, \bm{z}_2,...$ with their corresponding labels $l_1, l_2,...$ to train a logistic regression classifier. As all the weights $a^{(i)}_k$ are equal in \autoref{eq:aggregation}, the algorithm is not sensitive to the addition of a characteristic outlier in the bag. This drawback will be corrected in the next section thanks to the attention mechanism.

\subsection{Attention-Based Deep MIL models}
\label{subsec:abdmil}
We now describe an attention-based deep MIL method (which we will refer to as the \textit{deep MIL} method henceforth) that is an enhanced version of the model first proposed in \cite{ilse2018attention}. In this formulation, embeddings are pooled by taking a weighted average of the feature vectors. These weights correspond to learned parameters of a shallow neural network, named an attention mechanism for its ability to devote more ``attention" to important instances.

The setup and initial procedure is the same as described in \autoref{subsec:logistic}, except that we allow the coefficients $a^{(i)}_k$ to have different values defined by

\begin{equation}
\label{eq:attention}
a^{(i)}_k = \frac{\exp(\bm{w}^T\tanh(\bm{V}\bm{h}^{(i)}_k))}{\sum_{j=1}^n\exp(\bm{w}^T\tanh(\bm{V}\bm{h}^{(i)}_j))},
\end{equation}

\noindent where $\bm{w}$ and $\bm{V}$ are a parameterized vector and matrix respectively, and $\tanh$ is introduced to ensure a nonlinearity that includes both negative and positive values for proper gradient flow. Another difference to the procedure in \autoref{subsec:logistic} is that we multiply the attention values with the feature vector in a higher-dimensional feature space, notably \autoref{eq:aggregation} becomes

\begin{equation}
\label{eq:deepmil}
    \bm{z}_i = \sum_{k=1}^n a^{(i)}_k\tanh(\bm{V}\bm{h}^{(i)}_k) \, .    
\end{equation}

This idea of using the higher-dimensional feature space in the aggregation step is not always a common practice in multiple instance learning (in fact it is not used in \cite{ilse2018attention}) which is why we describe it as an enhanced version of the original method. The use of the higher-dimensional feature space in the aggregation steps is inspired by similar concepts which have been used in attention-based neural networks and sequence modeling, where transformed features are used to enhance representation power \cite{pmlr-v37-xuc15,https://doi.org/10.48550/arxiv.1706.03762,pmlr-v97-lee19d} which is the motivation to use it in this model. We demonstrate the advantage of using the higher-dimensional features compared with the original feature vector in \autoref{sec:lowdimmil}.

The final difference between the deep MIL model and the logistic regression model is that we use a feedforward neural network as a classifier instead of a logistic regression. By doing so we can use a single cost function in a unified framework, such that the gradient descent is performed both on the attention part (aggregation) and on the classification part. The parameters in $\bm{w}$ and $\bm{V}$ as well as the ones in the feedforward neural network are hence all trained together using the standard back-propagation methods involved in training a neural network.

In this work, we use the attention-based deep MIL model described above as well as the so-called ``gated" version (which we refer to as the \textit{gated deep MIL} model henceforth), which involves an additional sigmoid term leading to the equation for the coefficients to be
\begin{equation}
\label{eq:gatedattention}
a^{(i)}_k = \frac{\exp(\bm{w}^T\tanh(\bm{V}\bm{h}^{(i)}_k) \odot \textrm{sigm}(\bm{U}\bm{h}^{(i)}_k))}{\sum_{j=1}^n\exp(\bm{w}^T\tanh(\bm{V}\bm{h}^{(i)}_j) \odot \textrm{sigm}(\bm{V}\bm{h}^{(i)}_j))}
\end{equation}

and the equation for the aggregation to be

\begin{equation}
\label{eq:aggregationgated}
    \bm{z}_i = \sum_{k=1}^n a^{(i)}_k\tanh(\bm{V}\bm{h}^{(i)}_k)\odot\textrm{sigm}(\bm{U}\bm{h}^{(i)}_k) \, ,
\end{equation}

where $\odot$ represents element-wise multiplication. The original motivation for this additional term was the concern that $\tanh(x)$ is approximately linear for $x \in [-1, 1]$, which could limit the final expressiveness of learned relations among instances \cite{ilse2018attention}.

\subsection{The Attention-Based Linear MIL model}
\label{subsec:ablmil}
We also investigate a linear multiple instance learning model (which we refer to as the \textit{linear MIL} model henceforth), which is the same as the deep MIL model up to lack of the $\tanh$ nonlinearity, that is with the attention equation modified to

\begin{equation}
\label{eq:linearattention}
a^{(i)}_k = \frac{\exp(\bm{w}^T\bm{V}\bm{h}^{(i)}_k)}{\sum_{j=1}^n\exp(\bm{w}^T\bm{V}\bm{h}^{(i)}_j)}
\end{equation}
leading to the aggregation equation
\begin{equation}
\label{eq:linearaggregation}
    \bm{z}_i = \sum_{k=1}^n a^{(i)}_k\bm{V}\bm{h}^{(i)}_k \, .
\end{equation}

The motivation for investigating this model is to evaluate the necessity of the nonlinear components in the other two deep MIL models.

\subsection{The Attention-Based Extreme MIL Model}
\label{sec:abemil}

In order to improve on attention-based deep MIL, in particular with regard to the number of trained parameters, we introduce extreme learning machines and we will combine them with the attention-based MIL models in order to create a new MIL model. An introduction to the concept of extreme learning machines \cite{huang2006extreme} can be found in \autoref{sec:celm}.

The setup and procedure for an attention-based extreme MIL model (which we refer to as the \textit{extreme MIL} model henceforth) is identical to the attention-based deep MIL model described in \autoref{subsec:abdmil}, with the crucial difference that most parameters in \autoref{eq:attention} are now randomly initialized and fixed without training. The only parameters that are trained are the weights in $\bm{w}$. Despite these fixed parameters, this model still provides a universal function approximator (the reader can find more details in \autoref{sec:celm}). Therefore, it provides, in principle, an equivalent approach that requires much less compute power since many fewer parameters are trained. We will compare in \autoref{sec:results} the extreme MIL model to the deep MIL model as a benchmark: if extreme MIL offers comparable model performance, it will be more favorable than deep MIL as it uses by construction far fewer trained parameters, potentially also enhancing generalizability.

We present in \autoref{fig:elm_pipeline} our pipeline (upper part), in which the different variants for the attention-based MIL pooling are detailed in the lower part of the figure. In the extreme MIL variant, only the logistic regression parameters $\beta_i$ are trained, the weights vectors $\mathbf{w}_i$ are fixed.

To summarize, we will compare and benchmark the following five MIL models:
\begin{itemize}
    \item \textbf{Logistic regression} - The baseline using simple averaging to pool feature vectors and a logistic regression for classification.
    \item \textbf{Deep MIL} - The attention-based deep MIL model based on \autoref{eq:attention} and \autoref{eq:deepmil} for the attention mechanism, as well as using a feedforward neural network for the classification.
    \item \textbf{Gated deep MIL} - The gated version of the attention-based deep MIL model using \autoref{eq:gatedattention} and \autoref{eq:aggregationgated} for the attention mechanism.
    \item \textbf{Linear MIL} - The attention-based linear MIL model, where the $\tanh$ function is removed, based on \autoref{eq:linearattention} and \autoref{eq:linearaggregation}.
    \item \textbf{Extreme MIL} - the attention-based extreme MIL model, which, similar to the deep MIL model, is also based on \autoref{eq:attention} and \autoref{eq:deepmil} for the attention mechanism with the difference that only the final layer of parameters is optimized during the training process.
\end{itemize}

%
%

\section{Data Pre-processing}
\label{sec:datapreprocessing}

We use the BloodMNIST dataset to empirically compare these model architectures. In this section, we explain how we obtain the data and how we pre-process it to generate bags of feature vectors.

\begin{figure}[!ht]
    \centering
    \begin{subfigure}{0.60\textwidth}
        \centering
        \includegraphics[width=\linewidth]{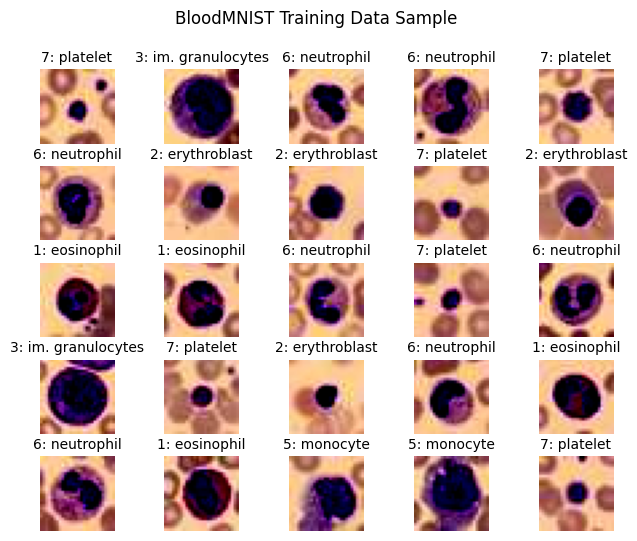}
    \end{subfigure}
    \vfill\vfill
    \begin{subfigure}{0.55\textwidth}
        \centering
        \includegraphics[width=\linewidth]{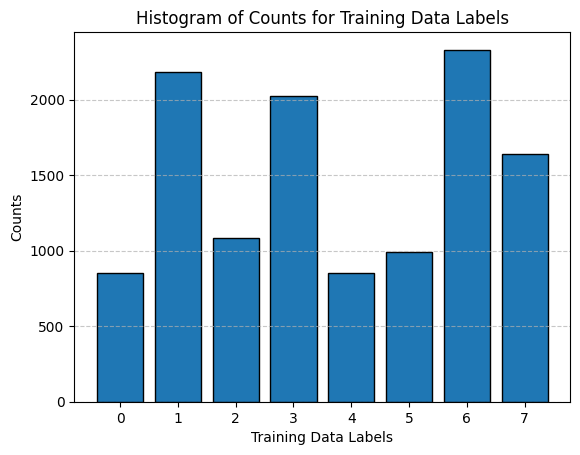}
    \end{subfigure}
    \caption{The Blood-MNIST training dataset. The top figure shows a sample of some of the individual training image data along with their class labels. The bottom figure is the histogram of the training dataset class labels.}
    \label{fig:training_data}
\end{figure}

The BloodMNIST data originates from the MedMNIST database \cite{medmnistv2} and contains 11,959 training images, 1,712 validation images, and 3,421 test images. For the purpose of this paper, we only use the training images (except when training a fine-tuned ResNet-18 model with individual images as described below and in \autoref{sec:finetuneresnet}). Each image has a label ranging from 0 to 7 indicating what type of blood cell it is (further biological explanations are further detailed in \cite{medmnistv2}). A sample of the training data along with the histogram of the label count can be found in \autoref{fig:training_data}.

The goal of the pre-processing step is to create 1000 training bags to use to tune the parameters of our different models and 1000 validation bags to evaluate the trained models and tune the hyperparameters of our models. To test the robustness of our model we create several permutations of these bags to train and validate our model on. Here we describe the process to make one such permutation.

We first split our 11,959 training images into a training dataset and a validation dataset using an 80–20 random split. The next step is to standardize the data and extract feature vectors from the images using a ResNET-18 neural network \cite{he2015deepresiduallearningimage}. While domain-specific feature extractors are still the gold standard for biomedical data, the growing adoption of multi-purpose models and foundation models must be acknowledged. Therefore, we apply two different approaches to training the ResNET-18: 1) where the ResNET-18 is pre-trained only on images from ImageNet \cite{Deng2009} (which we refer to as \textit{generic pre-processing} henceforth) and 2) where the ResNET-18 is fine-tuned using the same BloodMNIST data to generate a domain-specific model (which we refer to as \textit{specialized pre-processing} henceforth). A specialized pre-processing should in principle deliver better results as the pre-processing step is calibrated to the problem at hand, but is also by definition less generalizable than a generic pre-processing.  Building a generic model for multiple downstream applications (in a way, a foundation model) is a reality in many biomedical applications because data is so scarce. Furthermore, when you consider that microscopic data can be 10x, 20x, 40x, 50x, 100x magnification, it's clearly valuable to have models that generalize from one scale (ImageNet) to another (BloodMNIST). Comparing the outcome of the two methods will help us assessing the robustness of our entire pipeline and how much improvement the proposed methods have over traditional MIL methods for both cases. The details of the specialized pre-processing are described in \autoref{sec:finetuneresnet}.

\begin{figure}[!ht]
    \centering
    \begin{subfigure}[b]{\textwidth}
        \centering
        \includegraphics[width=0.8\textwidth]{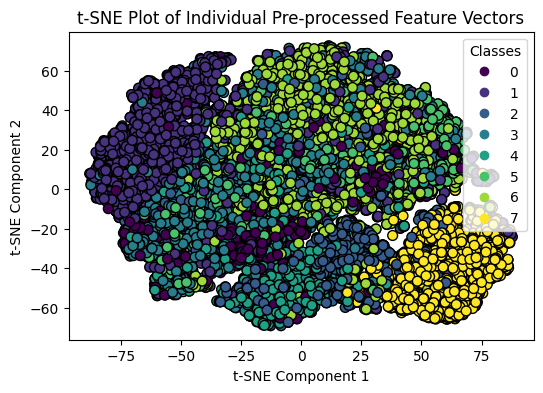} 
    \end{subfigure}
    \vspace{1em} 

    \begin{subfigure}[b]{\textwidth}
        \centering
        \includegraphics[width=0.8\textwidth]{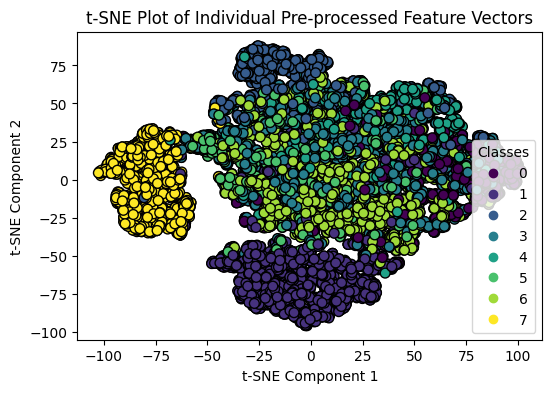} 
    \end{subfigure}

    \caption{Plot showing the first and second t-SNE components of the individually pre-processed training feature vectors after they have gone through a TSNE transformation. The pre-processing consists of passing each individual cell image through a ResNET-18 network pre-trained on ImageNet data (top) and pre-trained on the BloodMNIST data (bottom) and then taking the 8 first principle components of the output vectors.}
    \label{fig:tsne}
\end{figure}

These networks output feature vectors with 1,000 features each. We then perform a principle component analysis (PCA) \cite{Pearson1901,Hotelling1936} to reduce the number of features to 8. Following that, we perform the same steps to create the validation feature vectors using the validation dataset; however, we use the mean and standard deviation of the training set when standardizing and use the principle components computed on the training set when narrowing the number of features from 1,000 to 8. A t-SNE visualization of the data at this stage of the pre-processing can be seen in \autoref{fig:tsne}. As expected, the class representing erythroblasts (class 2) is to some extent more easily resolved in the t-SNE visualization for the ResNet-18 that was fine-tuned on domain-specific data.

We then create 1,000 training and 1,000 validation bags using the individual training and validation feature vectors. We train and evaluate our models on bag sizes varying from 1 to 30 with 20 different permutations for each bag size, but to simplify the description of the process, we will use a bag size of 10 as an example. This means that in each bag, there are 10 feature vectors. The goal is to train the model to recognize if at least one feature vector in a bag represents an erythroblast cell (class 2) or not. To balance the dataset, we create 500 ``negative'' bags that do not contain any feature vectors representing erythroblasts and 500 ``positive'' bags that contain exactly 1 feature vector representing an erythroblast and 9 that represent other cells. For each bag, we sample randomly with replacement across all available feature vectors in the set. Across all bag sizes, we ensure the 500 positive bags each contain 1 feature vector representing an erythroblast.

\begin{figure}[!ht]
    \centering
    \includegraphics[width=0.8\textwidth]{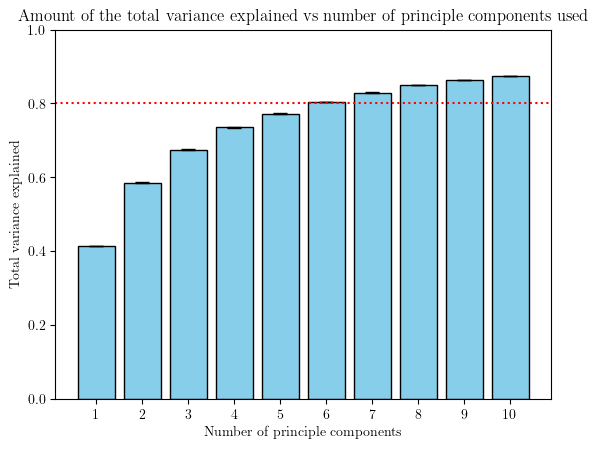}
    \caption{Plot showing the cumulative variance explained by principal components derived from the ResNet-18 feature vectors (trained on ImageNet) of the Blood-MNIST images. The values and their standard deviations (indicated by the barely visible error bars) have been computed by taking the principle components using 20 different random permutations of the training dataset. A dotted line has been drawn at 0.8 to guide the eye.}
    \label{fig:pca}
\end{figure}

A few comments concerning our pre-processing pipeline are in order:
\begin{itemize}
    \item At every step of the pipeline, we ensure that there is no information leakage from the validation set when training the models, including during the standardization and PCA processes. In both processes, the standardization and PCA parameters are calculated only using the training set, and then applied as-is to both the training and validation sets.
    \item We test two different pre-processing scenarios by using two different ResNET-18 networks (generic and specialized) to perform feature extraction on the single-cell images.
    \item We chose to use 8 principle components as our final feature vector as there are 8 classes in the BloodMNIST dataset. In addition we demonstrate that the first 8 principle components capture somewhere between 80–90\% of the total variance of the data as seen in \autoref{fig:pca} and that a leveling off in the vicinity of these 8 first principal components is also observed, giving us more grounds on the choice of these 8 principle components for the feature vector.
\end{itemize}

%
%

\section{Results}
\label{sec:results}
In this section, we explore the effect of attention mechanism architecture on detection of rare cell types in the blood. We look at bag sizes (collections of single-cell images) ranging from 1 to 30 feature vectors, where for each bag size we perform an evaluation using the process described in \autoref{sec:datapreprocessing}. In \autoref{sec:deepmilresults}, we use the simple averaging and logistic regression described in \autoref{subsec:logistic} as a baseline against which we compare the deep MIL and gated deep MIL models. In \autoref{sec:linearmilresults}, we compare the linear MIL model to the previous deep MIL model to study the effect that the nonlinear components of the network has on its performance. Finally in \autoref{sec:emilresults}, we compare the extreme MIL model to the deep MIL model to evaluate the effect of freezing most of the parameters in the attention network. All models were subjected to same random train-validation splits in a consistent manner throughout the whole process. In addition, the hyperparameters of the models were optimized by doing a grid search, of which more details can be found in \autoref{sec:hyperparameters}, which also includes a discussion of the ineffectiveness of using the early stopping method. Statistical significance between the model performance curves was evaluated using a modified Chi-squared method with a correction for small sample sizes and a Bonferroni correction for multiple comparisons across the curves \cite{hristova2023statsigcurves} 

\subsection{Deep Multiple Instance Learning Benchmarks}
\label{sec:deepmilresults}

\begin{figure}[h!]
    \centering
    \begin{subfigure}{0.49\textwidth}
        \centering
        \includegraphics[width=\linewidth]{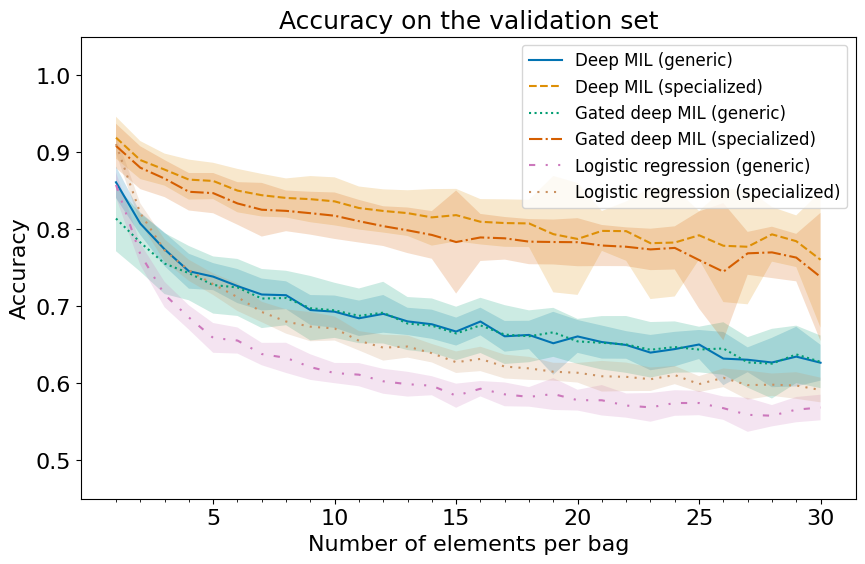}
    \end{subfigure}
    \hfill
    \begin{subfigure}{0.49\textwidth}
        \centering
        \includegraphics[width=\linewidth]{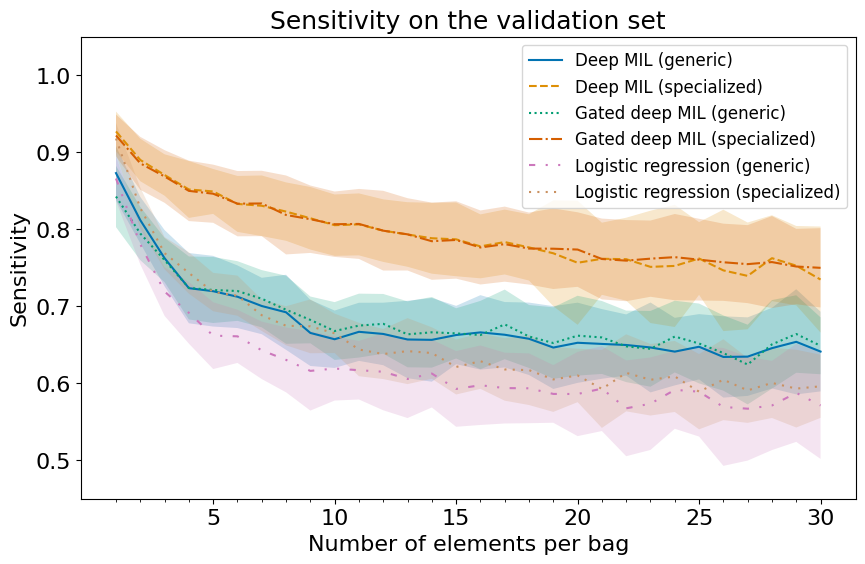}
    \end{subfigure}

    \vspace{0.1cm} 

    \begin{subfigure}{0.49\textwidth}
        \centering
        \includegraphics[width=\linewidth]{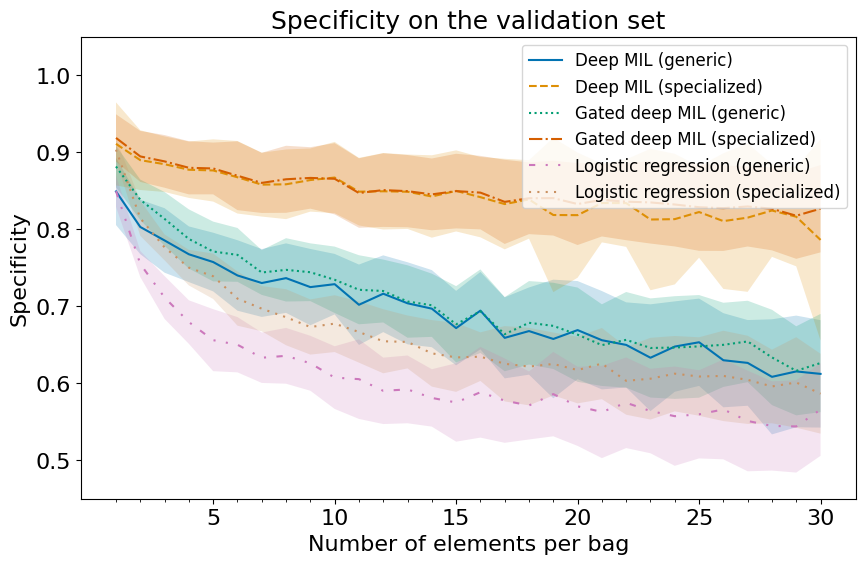}
    \end{subfigure}
    \hfill
    \begin{subfigure}{0.49\textwidth}
        \centering
        \includegraphics[width=\linewidth]{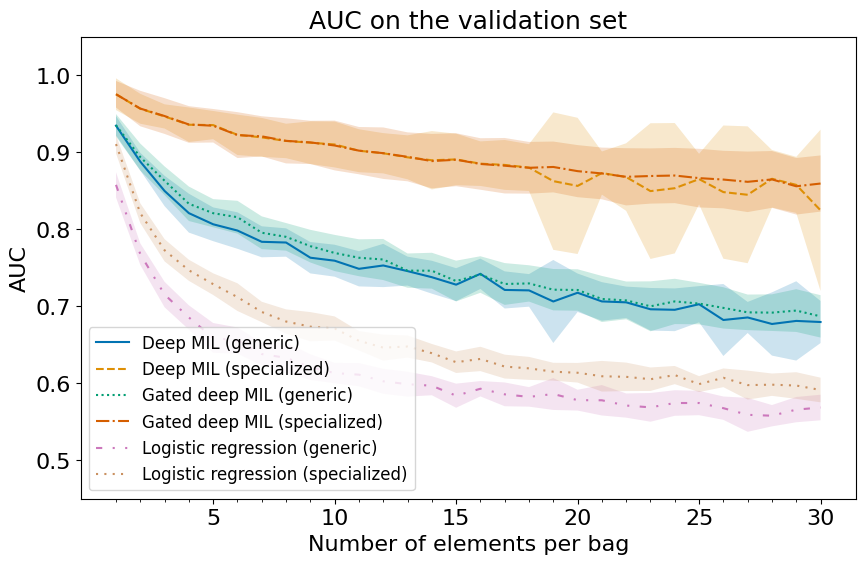}
    \end{subfigure}

    \caption{Comparison of deep and gated deep MIL models to the logistic regression baseline model. The accuracy, sensitivity, specificity, and AUC are measured for models trained on either natural imagery (``generic") or domain-specific (``specialized") data. The lines and bands for all models represent the calculated means and standard deviations respectively over 20 different random permutations of the train-validation split.}
    \label{fig:deepmil}
\end{figure}

In this section, we investigate the performance of the deep MIL and gated deep MIL models trained on either the generic or domain-specialized data and compare their performance to the logistic regression model. The performance of these models can be seen in \autoref{fig:deepmil}, where we evaluate the accuracy, sensitivity, specificity, and area under the curve (AUC) of the models on the data. The logistic regression baseline model exhibited the poorest performance across all four metrics, regardless of whether it was trained on generic or specialized data. Compared to logistic regression, deep MIL and gated deep MIL models offered performance improvements of 15-25\% on average across all metrics. In addition, the image domain on which these models were trained had a large impact on performance. Specialized deep MIL and gated deep MIL models outperformed their generic counterparts by 10-15\% in all metrics, with the generic models falling below 0.80 AUC above 7 cells per bag. Interestingly, no significant difference was observed between the gated deep MIL model compared to its deep MIL counterpart (\autoref{tab:fig6_stats}).

These results suggest that adoption of a validated attention mechanism architecture in a new imaging domain may not always yield the expected performance benefits. Even so, the attention mechanisms in both deep MIL and gated deep MIL resulted in a demonstrable benefit over simple weighted averaging, when applied to rare cell detection. In addition to our main results above, we also demonstrate the effectiveness of increasing the higher-dimensional feature space in the aggregation step which is discussed more in detail in \autoref{sec:lowdimmil}.

\subsection{Linear Multiple Instance Learning Benchmarks}
\label{sec:linearmilresults}

\begin{figure}[h!]
    \centering
    \begin{subfigure}{0.49\textwidth}
        \centering
        \includegraphics[width=\linewidth]{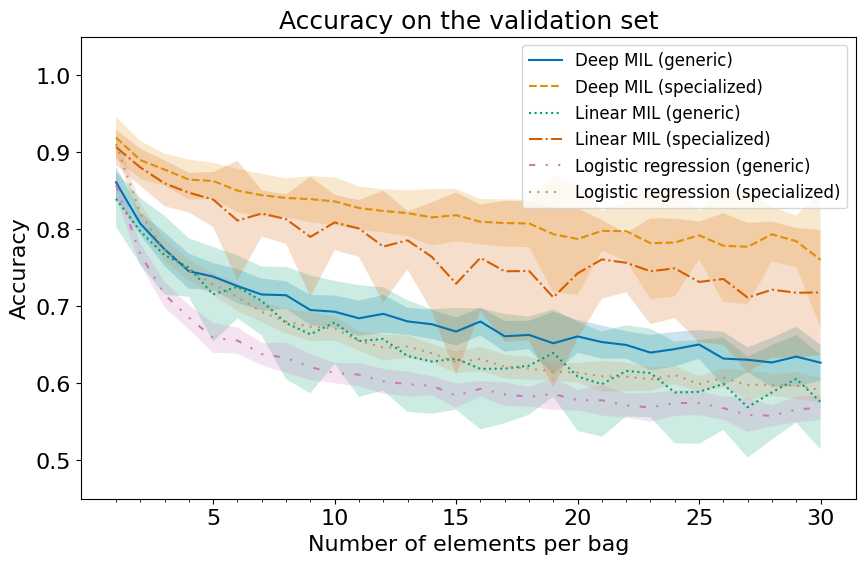}
    \end{subfigure}
    \hfill
    \begin{subfigure}{0.49\textwidth}
        \centering
        \includegraphics[width=\linewidth]{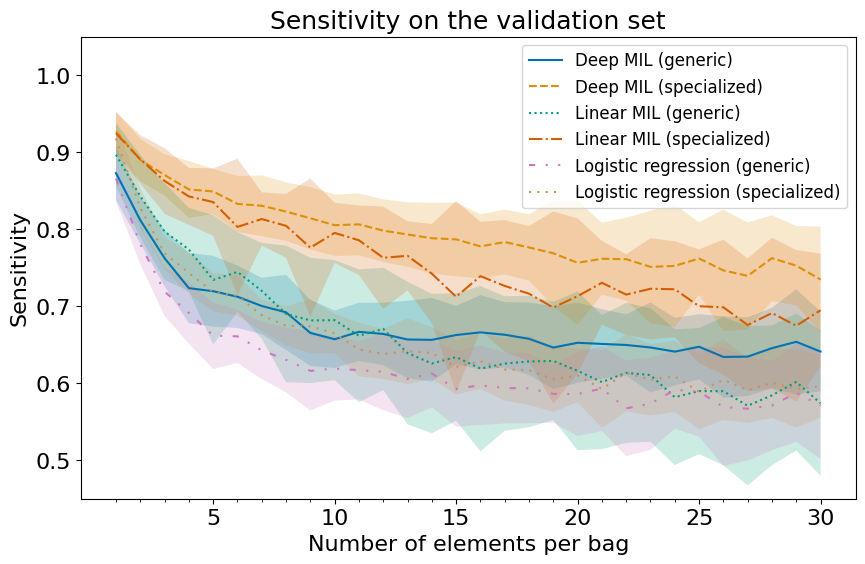}
    \end{subfigure}

    \vspace{0.1cm} 

    \begin{subfigure}{0.49\textwidth}
        \centering
        \includegraphics[width=\linewidth]{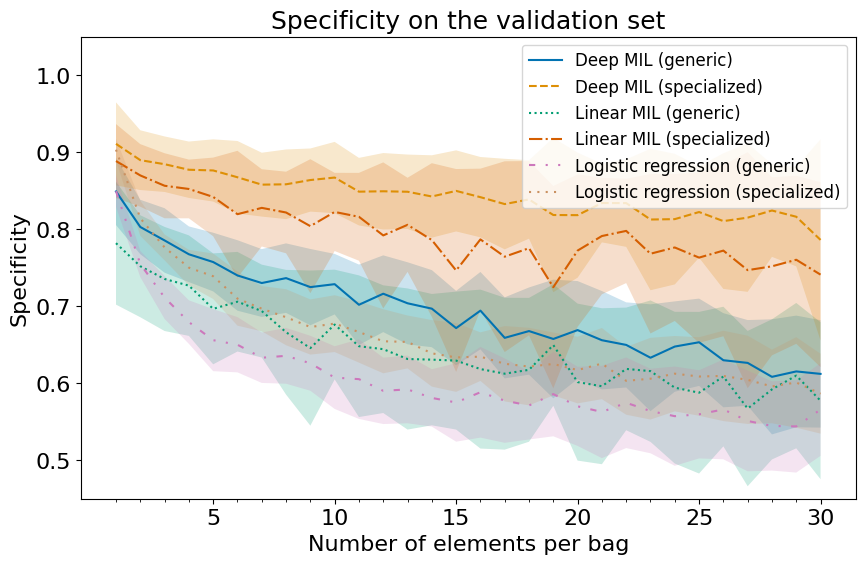}
    \end{subfigure}
    \hfill
    \begin{subfigure}{0.49\textwidth}
        \centering
        \includegraphics[width=\linewidth]{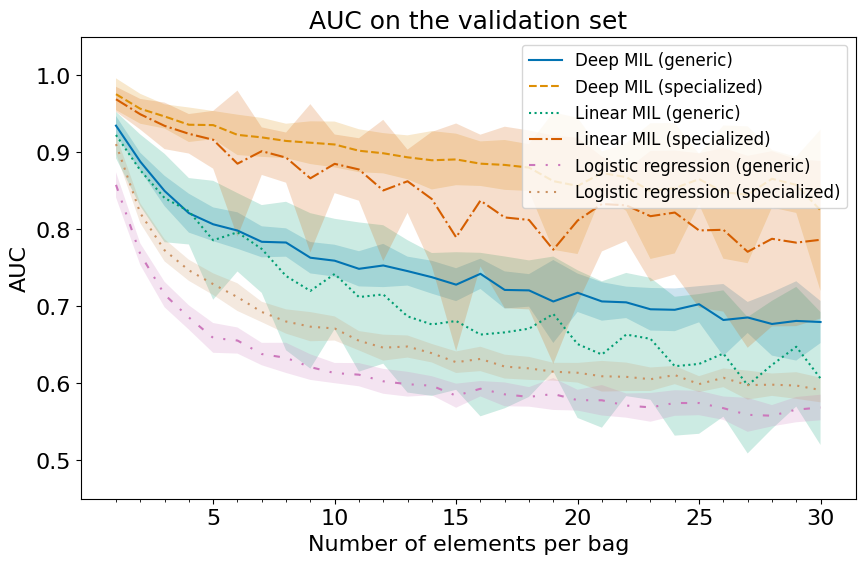}
    \end{subfigure}

    \caption{Comparison of the linear MIL and deep MIL models to the logistic regression baseline model. The accuracy, sensitivity, specificity, and AUC are measured for models trained on either natural imagery (``generic") or domain-specific (``specialized") data. The lines and bands for all models represent the calculated means and standard deviations respectively over the permutations of the bags over 20 different random permutations of the train-validation split.}
    \label{fig:linearmil}
\end{figure}

In this section, we evaluate the performance of the linear MIL model on the generic pre-processed and specialized pre-processed data to validate the role of a nonlinear activation function in the attention mechanism. Having established that there is no significant difference in the performance of the deep MIL and gated deep MIL models, we omit the results of the gated deep MIL models in the plots for improved clarity. The performance of these models can be seen in \autoref{fig:linearmil}, where we evaluate the accuracy, sensitivity, specificity, and area under the curve (AUC) of the models on the data. Over all bag sizes and performance metrics, linear MIL outperformed its counterpart logistic regression model. Training on domain-specialized data once again significantly improved model performance across all metrics for all architectures. Omission of the nonlinearity in the attention mechanism, transforming deep MIL into a linear MIL architecture, resulted in a modest but significant decrease in all performance metrics of about 4-5\% (\autoref{tab:fig7_stats}). More pronounced was the effect on the model stability: an F-test for equality of variances revealed that accuracy of the deep MIL models maintained a tighter confidence interval over the 20 train-validation splits than did the linear MIL counterparts (p<0.01 and effect size greater than 2.5 for both generic and specialized versions).

We also explore a scenario where we decrease the stability of the deep MIL models by introducing early stopping and attempt to increase the stability of the linear model by allowing more epochs to converge to allow the linear MIL model as large an advantage as possible in terms of stability. Even in this scenario, we find the linear MIL to be at best on par with the unstable deep MIL models (\autoref{sec:hyperparameters}).

Taken together, these results recapitulate the validity of an attention-based approach for rare event detection and highlight the importance of nonlinearities to build a high-performing and stable model. 

\subsection{Attention-based Extreme Multiple Instance Learning Benchmark}
\label{sec:emilresults}
\begin{figure}[h!]
    \centering
    \begin{subfigure}{0.49\textwidth}
        \centering
        \includegraphics[width=\linewidth]{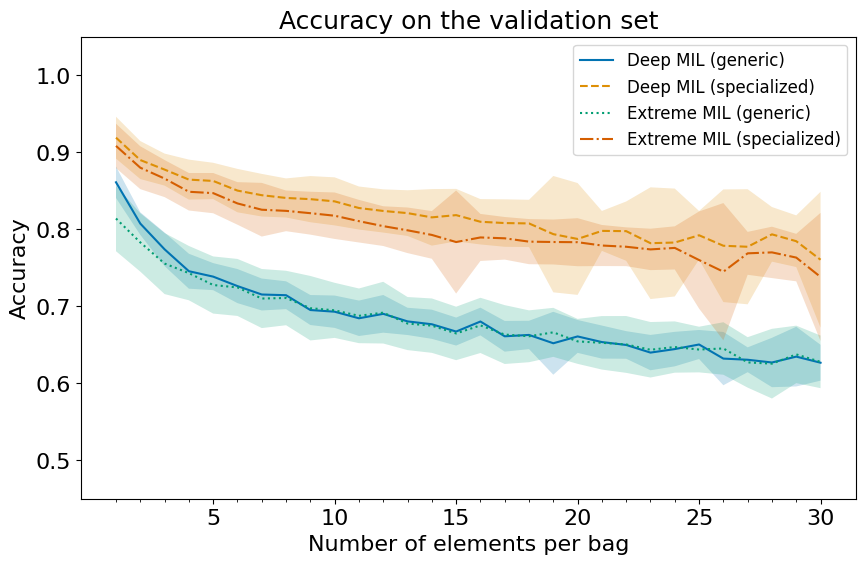}
    \end{subfigure}
    \hfill
    \begin{subfigure}{0.49\textwidth}
        \centering
        \includegraphics[width=\linewidth]{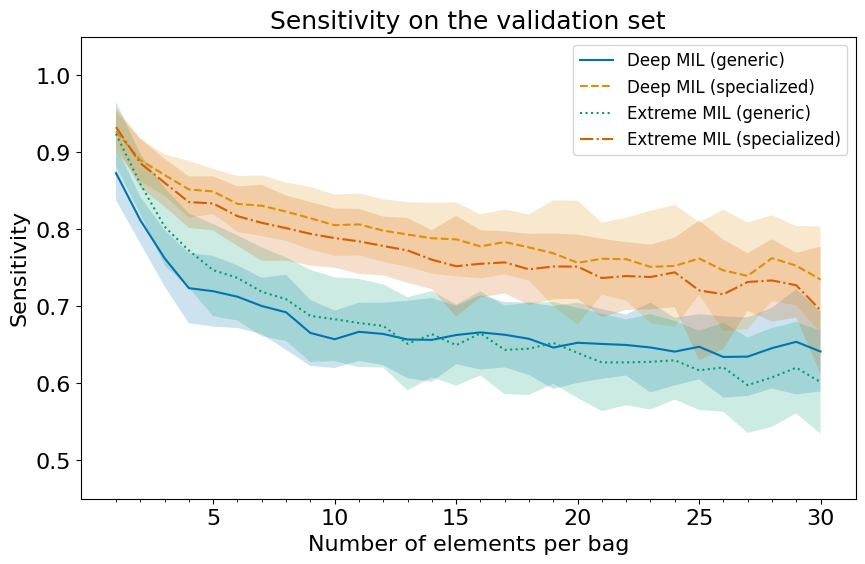}
    \end{subfigure}

    \vspace{0.1cm} 

    \begin{subfigure}{0.49\textwidth}
        \centering
        \includegraphics[width=\linewidth]{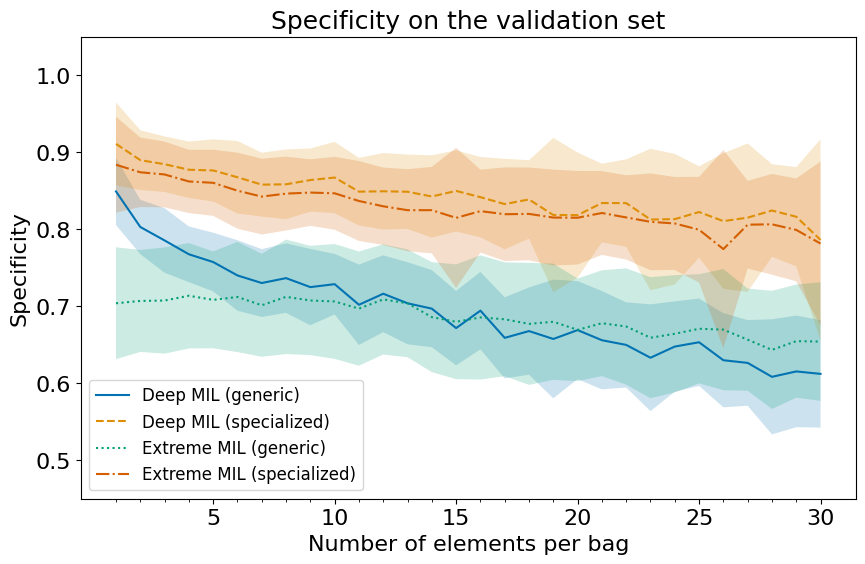}
    \end{subfigure}
    \hfill
    \begin{subfigure}{0.49\textwidth}
        \centering
        \includegraphics[width=\linewidth]{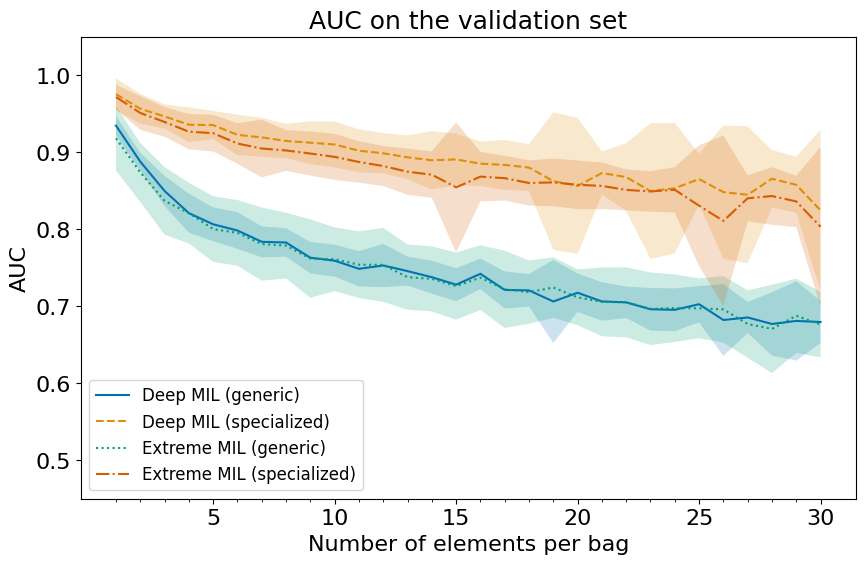}
    \end{subfigure}

    \caption{Comparison of the extreme MIL model to the deep MIL model. The accuracy, sensitivity, specificity, and AUC are measured for models trained on either natural imagery (``generic") or domain-specific (``specialized") data. The lines and bands for all models represent the calculated means and standard deviations respectively over the permutations of the bags over 20 different random permutations of the train-validation split.}
    \label{fig:emil}
\end{figure}

In this section we evaluate the performance of the extreme MIL model on the generic pre-processed and specialized pre-processed data to investigate the effect of drastically decreasing the number of trained parameters in the model. As in the previous section, we limit the plots of this model to the deep MIL and extreme MIL models for improved clarity. The performance of these models can be seen in \autoref{fig:emil}, where we evaluate the accuracy, sensitivity, specificity, and area under the curve (AUC) of the models on the data. Despite the untrained parameters in the extreme MIL attention mechanism, these models significantly outperformed logistic regression across all metrics, boosting AUC by 21.8\% in the specialized version. As seen for previous attention-based architectures, training on domain-specialized data resulted in a boost of 10-14\% across all metrics. Although the specialized deep MIL model significantly outperformed its extreme MIL counterpart for all metrics but specificity, the improvements were limited to just 1-2\%. Therefore, the extreme MIL architecture offered substantially similar accuracy and AUC as the deep MIL, but with fewer trained parameters. This surprising result suggests that the use of an attention mechanism and training on domain-specialized data are far more important factors than the number of trained parameters.

%
%

\section{Discussion and Outlook}
\label{sec:discussion}

These structured experiments illustrate the importance of model architecture when training a diagnostic model to detect rare cell types. In \autoref{fig:deepmil}-\autoref{fig:emil}, the performance of each model decayed as the bag size increased. Since each bag was composed of at most one erythroblast, these plots can be interpreted as the ability of the model to detect erythroblasts at lower concentrations, with the plateau representing the detection limit. The plateau of the model performance curves as bag size grows is especially relevant for the application of erythroblast detection. Clinical studies report erythroblast levels of 1-200 per microliter of blood as being relevant for prognosis of critically ill patients in intensive care units \cite{Stachon2007}. Assuming a white blood cell count of 4000 to 10,000 per microliter, the bag size must minimally exceed 20 cells in order to yield a single erythroblast in patients with leukopenia, with larger bag sizes minimally required for broader clinical application. All MIL-based models exhibited higher plateaus than their logistic regression counterparts, motivating the use of carefully designed attention mechanisms in this application.

The comparison of linear MIL, deep MIL, and gated MIL architectures demonstrate that nonlinearities can be powerful tools against model overfitting but only when applied in the right context. Regardless of whether the model was trained on generic or specialized imagery, the addition of a gate to the deep MIL attention mechanism (``gated deep MIL") did not significantly improve model performance or stability (\autoref{fig:deepmil}, \autoref{tab:fig6_stats}). But when stripped of its nonlinear activation function, the deep MIL model becomes a linear MIL architecture, dropping in performance by 4-5\% across all performance metrics (\autoref{tab:fig7_stats}). In addition to reducing expressivity of the model, loss of the nonlinearity in the linear MIL architecture also caused confidence intervals to double, suggesting a less stable model. While these results recapitulate the utility of nonlinear activation functions in attention mechanisms, they also underscore that empirical validation is a crucial step when applying an attention mechanism architecture to a new image domain. 

In all cases, the model trained on the specialized blood dataset outperformed its generic counterpart. This is not surprising, since blood cell images are imaged at 40-50x higher resolution than the natural imagery found in ImageNet. Interestingly, the performance of the linear MIL model over its logistic regression counterpart depended greatly on whether the models were trained on generic or specialized data. When trained on generic imagery, the linear MIL model only offered a 3-4\% improvement in accuracy, sensitivity, and specificity compared to the logistic regression. But when trained on specialized blood cell imagery, the gap tripled for a 10-13\% boost (\autoref{tab:fig7_stats}). We hypothesize that the expressiveness of a MIL approach with attention allows the model to better capture the nuances of the biomedical image data, particularly as the bag size grows. It remains to be seen whether this performance boost will extend to image data prepared in other labs.

Continuing our investigation of attention architecture, we turned to the number of parameters in the attention network. When increasing the number of hidden nodes in the attention mechanism from $2^5$ to $2^{11}$, we observed a contraction in the confidence intervals around $2^8$ nodes (\autoref{fig:numhiddennodes}). This suggests that a sizable hidden layer in the attention mechanism can be beneficial to model stability and performance. Furthermore, extreme MIL models trained on specialized imagery were substantially similar in performance to their deep MIL counterparts, conceding just 1-2\% performance in accuracy, sensitivity, and AUC while maintaining specificity (\autoref{fig:emil}, \autoref{tab:fig8_stats}). This counter-intuitive result demonstrates that model performance is not solely a function of the number of trained parameters and that a clinically useful model can be achieved with less training time and expense. An elaboration of this experiment can be found in \autoref{sec:hiddennodes}.

Though it is beyond the scope of this work, careful threshold selection and model generalization represent critical next steps in model development for erythroblast detection. A threshold that prioritizes sensitivity could allow for earlier detection of erythroblasts and intervention, possibly preventing death. The logistic regression models demonstrated robust accuracy and AUC across the 20 permutations of train-validation splits, while sensitivity and specificity varied more widely. The same phenomenon was observed for deep MIL, gated deep MIL, and extreme MIL models, though less pronounced. Variability in sensitivity and specificity suggests that integration of the model in different diagnostic laboratories could result in unpredictable shifts in clinical utility. Along with selection of the right model architecture, training data should comprise a multi-center cohort to ensure reproducible performance in a clinical setting.

\subsection{Alternative Attention-based Mechanisms}

We have proposed and studied two variants of deep MIL models (gated and non-gated) and have observed that nonlinearities play an important role in enhancing the quality of the model. Our goal is also to reduce the number of trained parameters and this is achieved with our proposed extreme MIL.

It should be nonetheless mentioned that there exist more advanced versions of the attention-based deep MIL method that effectively involve more complex nonlinearities. A study published in 2020~\cite{shi2020loss} introduced the concept of loss-based attention, using the loss function (softmax+cross-entropy) for calculating the instance attention weights. More recently, inspired by the transformer mechanism, reference~\cite{keshvarikhojasteh2024} introduced multi-head attention-based deep MIL in the context of whole-slide image classification in digital pathology. The core idea is to run the attention mechanism in parallel across multiple heads to capture diverse aspects of relationships. Competitive results were obtained on public dataset of lung and kidney images. All these ideas could also be combined with our proposed extreme MIL to potentially further enhance the results we have obtained.

\subsection{Quantum Extreme Learning Machines for the Attention-based Extreme MIL Model}
\label{subsec:qelm}

Given the benchmark results obtained when comparing the linear deep MIL with both the deep MIL and gated deep MIL, it is clear that nonlinearities in the network can be beneficial in improving the outcome of the deep MIL. The most common way to use quantum networks is to first encode the input into a quantum state which is then evolved via parametrized unitary operators after which the expectation values of observables are measured \cite{PhysRevA.103.032430}. These expectations values can be used either directly as a prediction or as an addition variable (e.g. an attention weight) for further calculations. The two main sources of nonlinearity come from the the encoding step, where the encoded state depends on the features in a nonlinear way, and the parametrized operators act on the quantum state in an nonlinear manner with respect to the parameters. Also, the quantum state space naturally provides a higher-dimensional feature space that can be taken advantage of, which, as we have seen, can significantly improve the results.

Moreover, the results obtained with our novel extreme MIL demonstrate comparable performance to the (gated) deep MIL, but using far fewer trained parameters. This is a very good starting point for a quantum extension of the model, as most of the quantum nodes replacing the classical nodes will use fixed (randomized) parameters and hence do not require training and multiple evaluation on a quantum system beyond the measurement mechanism. In addition, with quantum networks having the potential for a better expressivity compared to their classical counterparts, using them in the extreme machine learning context avoids the well-known problem of barren plateaus \cite{McClean2018}. It is thus natural to investigate the quantum version of an ELM in our novel algorithm presented in \autoref{sec:abemil}. 

The global architecture of the quantum version of an ELM, known as a quantum ELM (QELM) \cite{Mujal2021, Innocenti2023, Xiong2023}, is similar to that of a classical ELM. The difference comes from the nonlinear activation function which is replaced by a quantum circuit. In particular, the QELM can still be described by \autoref{eq:elm}, as outlined in \autoref{sec:celm}, but now the function $g$ is given as $g(\theta,\cdot)$ and encapsulates the output of a parameterized quantum circuit with parameters $\{\theta\}$. The first part of the quantum circuit is the data embedding layer where the data is stored in a quantum state $\ket{\psi(\bm{h}_j)}$, then this quantum state is evolved into a final quantum state $\ket{\psi(\theta, \bm{h}_j)}$ thanks to parameterized unitaries randomly initialized and fixed. The generic structure of $g(\theta,\bm{h}_j)$ is given as
\begin{equation}
    g(\theta,\bm{h}_j) = \bra{\psi(\theta,\bm{h}_j)} \sigma^{(i)} \ket{\psi(\theta,\bm{h}_j)},
    \label{eq:qelm}
\end{equation}
where $\sigma^{(i)}$ is a given Pauli operator acting on the $i$th qubit. A standard choice is the Pauli $Z$ operator $\sigma^z$. This framework can be in principle extended to any set of observables.

There are multiple directions to explore in order to improve our attention-based extreme MIL with a QELM instead of a classical ELM. In particular:

\begin{itemize}
\item Which encoding scheme for the classical data onto a quantum state is chosen? There exist multiple variants such as amplitude encoding, angle encoding, and hybrid schemes.
\item Which operators for the quantum measurement give the best outcome?
\item How does the application of a QELM on real quantum hardware differ from the simulation of the ideal case of noiseless quantum systems?
\end{itemize}

These ideas could be fruitful research avenues in future work in order to further enhance the robustness and usefulness of the attention-based extreme MIL algorithm described in the present paper.


\subsection{Outlook}
Whole-slide image classification is a problem of great interest in the biomedical sciences. On the one hand, effective solutions can quicken progress towards accurate and efficient diagnostics at the level of an individual. On the other hand, it encapsulates a range of challenges that also need to be addressed as advanced AI and quantum computing algorithms are applied to other areas in medicine, including handling large and small data sets, dealing with noisy data, and building robust yet explainable models. While a range of research problems must still be tackled, the present work can help accelerate research efforts in this space.


\bibliographystyle{splncs03_unsrt}
\bibliography{bibliography}
%
%
%
%

\appendix

\section{Extreme Learning Machines}
\label{sec:celm}
\begin{figure}[!h]%
\centering
\includegraphics[width=0.8\linewidth]{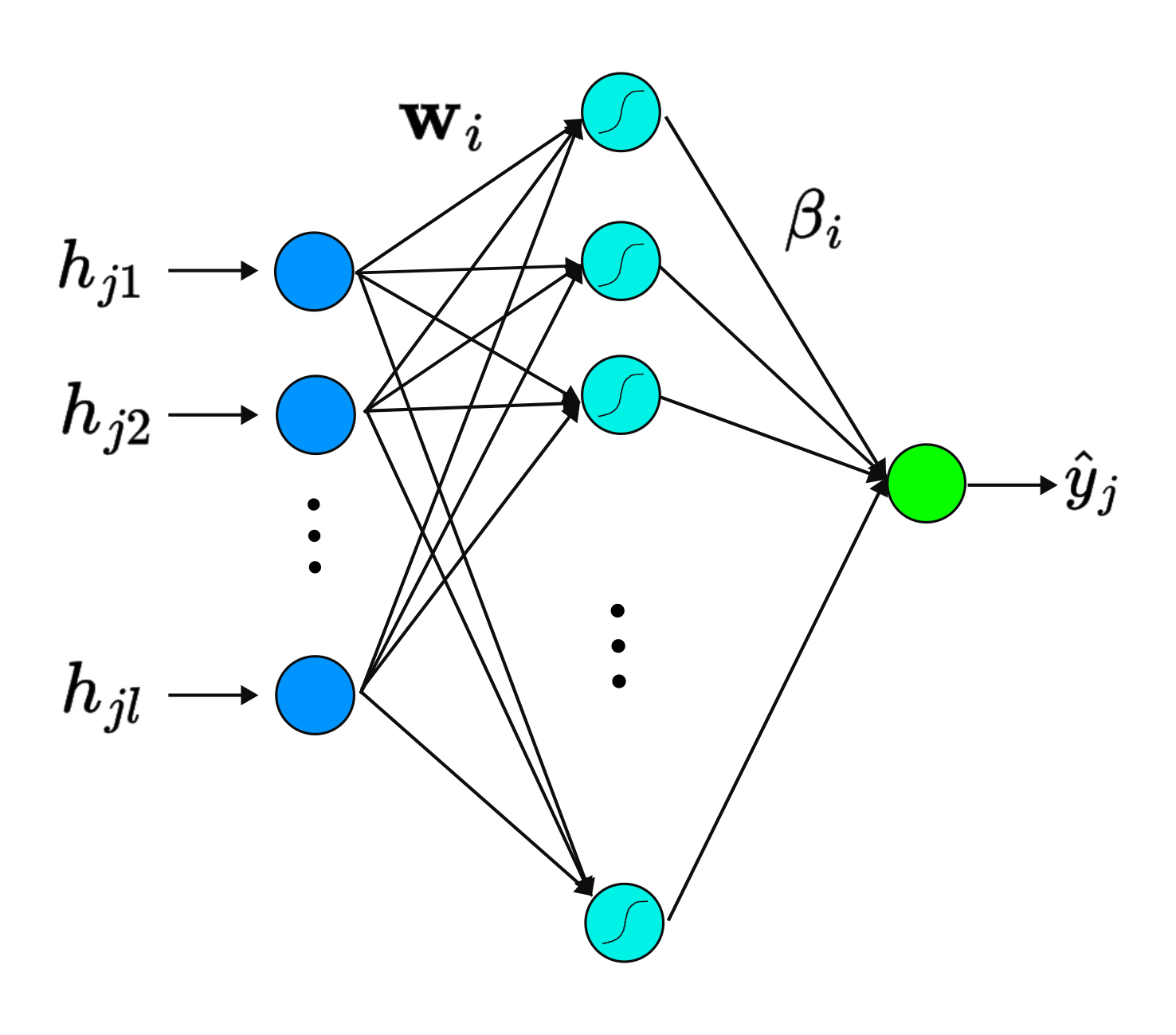}
\caption{A visual representation of the SLFN network used in typical ELMs. $h_{j1}, h_{j1},...,h_{jl}$ represent the features of the input vector, $\bm{h}_j$, $\bm{w}_i$ represents the randomized weights, $\beta_i$ the trainable parameters, and $\hat{y}_j$ the prediction. We show the nonlinear activation functions as sigmoid functions.}
\label{fig:elm}
\end{figure}

The extreme learning machine (ELM) model consists of a single-layered feedforward neural network (SLFN) followed by a linear regression. If we assume that our input data consists of vectors $\bm{h}_j$, the ELM model can be described as
\begin{equation}
    \hat{y}_j = \sum_{i=1}^m \beta_i g(\bm{w}_i^T\bm{h}_j),
    \label{eq:elm}
\end{equation}    

\noindent where $\hat{y}_j$ is the predicted output, $m$ is the number of nodes in the hidden layer, $\beta_i \in \mathbb{R}$, and $g(\cdot)$ is a nonlinear activation function such as a sigmoid function\footnote{In theory, the expression $g(\bm{w}^T\bm{h}_j)$ can be replaced by any nonlinear function that is dependent on $\bm{h}_j$, including a literal bucket of water \cite{fernando2003bucket}. Still, mostly when one uses the term ELM, a SLFN is implied.}. The particularity of the ELM is that the weights of the hidden layers, i.e. the vector $\bm{w}$, are initialized randomly at the start of the training and are then permanently fixed. Therefore, the only parameters that are trained are the $\beta_i$ parameters, which is done using a linear regression, typically by minimizing the mean squared error across the training data. A visual representation of the network used in ELMs can be seen in \autoref{fig:elm}.

Here we have described the simplest form of an ELM, and many modifications are possible. For example, variants of the simple ELM can include bias terms to make it more expressive or a regularization term to make it more robust, replacing $\beta_i$ with $\bm{\beta_i} \in \mathbb{R}^r$ if we want our outputs to be vectors of length $r$. We direct readers to \cite{wang2021review} for more details on the many possible modifications. In addition, ELMs algorithms are part of a larger family of algorithms that use randomized networks such as reservoir computing (RC) models and random vector functional link (RVFL) networks \cite{markowskakaczmar2021extreme}. However, for the purposes of this paper, we will refer to ELMs as the model described by \autoref{eq:elm} without any further modifications.

It may seem counter-intuitive that a randomized network can be used to approximate an arbitrary function without training all the parameters. Nevertheless, it has been shown that such a randomized network is a universal function approximator for continuous functions \cite{huang2006universal}\footnote{The outline of the proof in \cite{huang2006universal} is as follows: the authors start by defining any continuous function $f$ that they wish to approximate and then take a randomly generated SLFN such as the one shown in \autoref{fig:elm} with $m-1$ hidden nodes, where each node has a nonlinear bounded non-constant or integrable piecewise continuous activation function $g(\cdot)$. They define the approximation given by this network as $f_{m-1}$. They then prove that if another randomly generated node with weights $\bm{w}_m$ is added to the hidden layer, the approximation error of this new network is minimized if the value of the added coefficient is $\beta_m=\langle||f-f_{m-1}||,g_m\rangle/||g_m||^2$, where $g_m(\bm{h}_j)=g(\bm{w}_m^T\bm{h}_j)$, which represents the contribution of the newly added hidden node. They then prove that the sequence $||f-f_m||$ converges and finally that $\lim_{m\rightarrow \infty}||f-f_m||=0$.}. The intuition is that if the input vectors are projected into a set of new features in a sufficiently high-dimensional space via nonlinear functions, the linear combination of these new functions can approximate any continuous function, similar in spirit to how support vector machines with Gaussian kernels are also universal function approximators \cite{park1991universal,huang2014insight}.


\section{Hyperparameter selection of MIL models}
\label{sec:hyperparameters}
The main criteria in choosing the hyperparameters have been the shape and convergence of the loss curve, as well as quality of the AUC on the validation set and the stability of the sensitivity and specificity. For every hyperparameter, a grid search was performed where each time the model was evaluated on 3–5 different sets of random train/validation splits for 2, 5, 10, 20, and 30 bags per cell. We will first discuss the hyperparameter selection for the generic pre-processed data, followed by the hyperparameter selection for the specialized pre-processed data. In addition, when training all models the Adam optimizer \cite{kingma2015} and binary cross-entropy loss were used.

\begin{figure}[!ht]
    \centering
    \begin{subfigure}{0.45\textwidth}
        \centering
        \includegraphics[width=\linewidth]{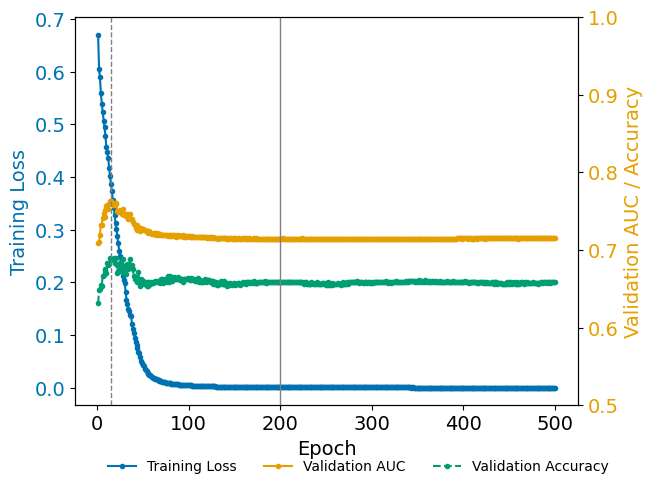}
    \end{subfigure}
    \hfill
    \begin{subfigure}{0.45\textwidth}
        \centering
        \includegraphics[width=\linewidth]{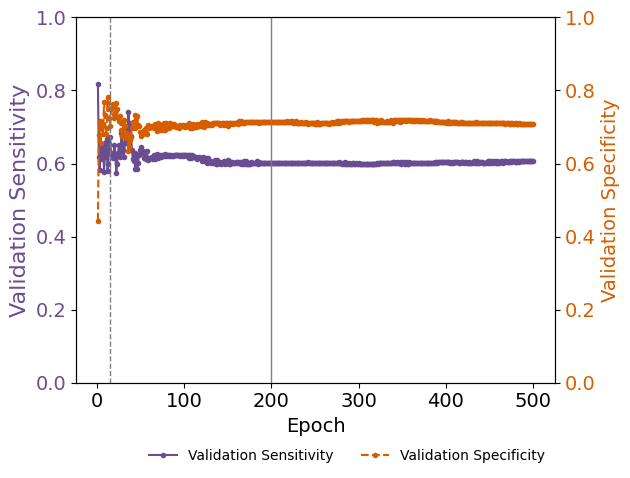}
    \end{subfigure}


    \begin{subfigure}{0.45\textwidth}
        \centering
        \includegraphics[width=\linewidth]{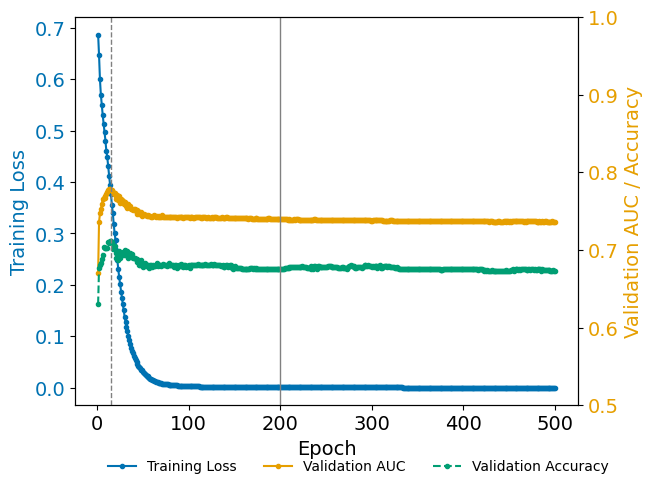}
    \end{subfigure}
    \hfill
    \begin{subfigure}{0.45\textwidth}
        \centering
        \includegraphics[width=\linewidth]{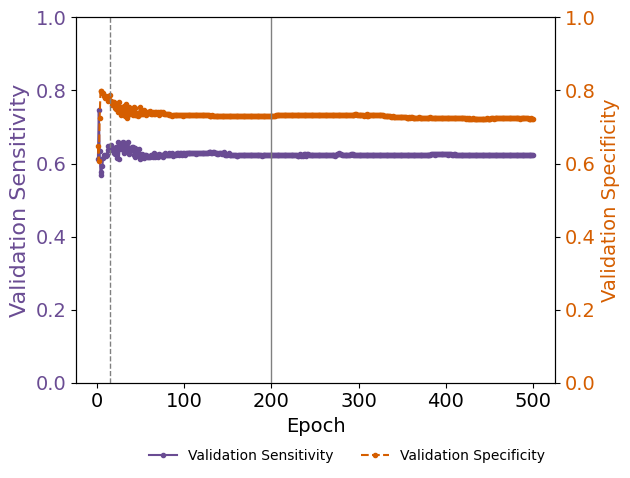}
    \end{subfigure}

    \begin{subfigure}{0.45\textwidth}
        \centering
        \includegraphics[width=\linewidth]{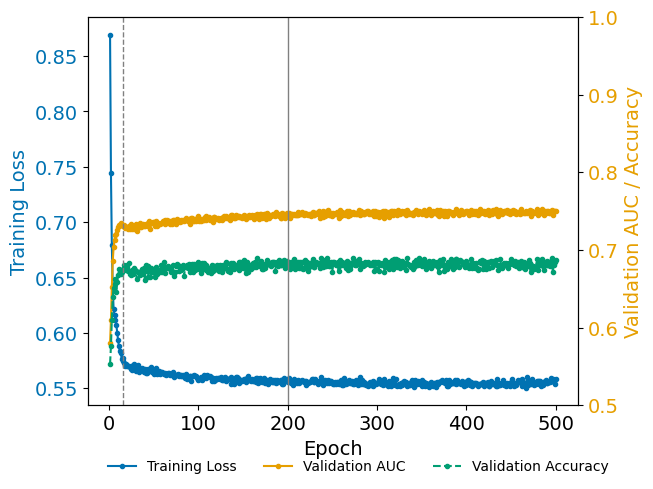}
    \end{subfigure}
    \hfill
    \begin{subfigure}{0.45\textwidth}
        \centering
        \includegraphics[width=\linewidth]{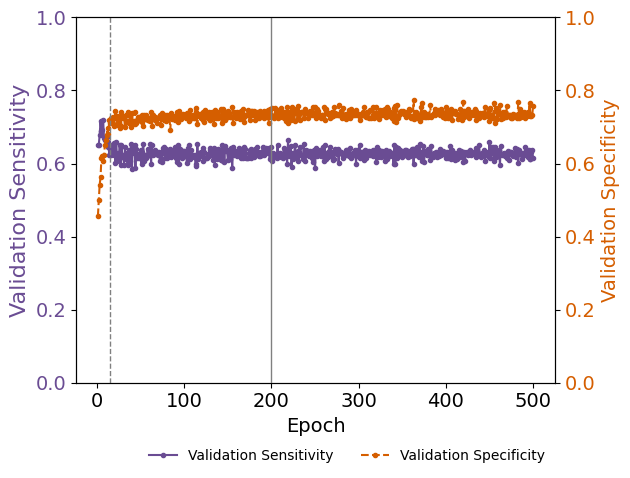}
    \end{subfigure}

    \begin{subfigure}{0.45\textwidth}
        \centering
        \includegraphics[width=\linewidth]{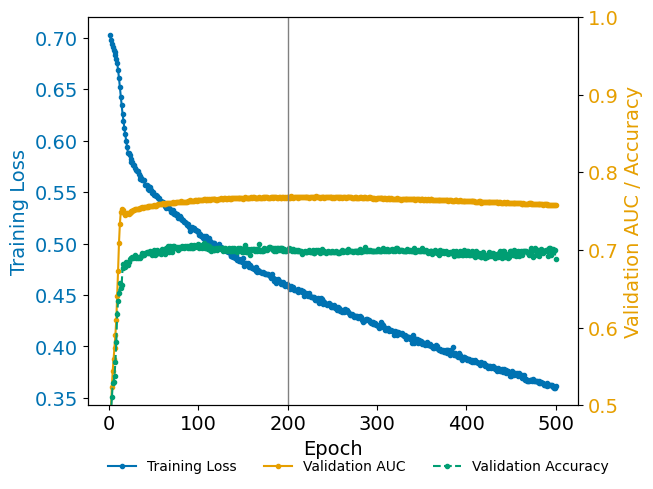}
    \end{subfigure}
    \hfill
    \begin{subfigure}{0.45\textwidth}
        \centering
        \includegraphics[width=\linewidth]{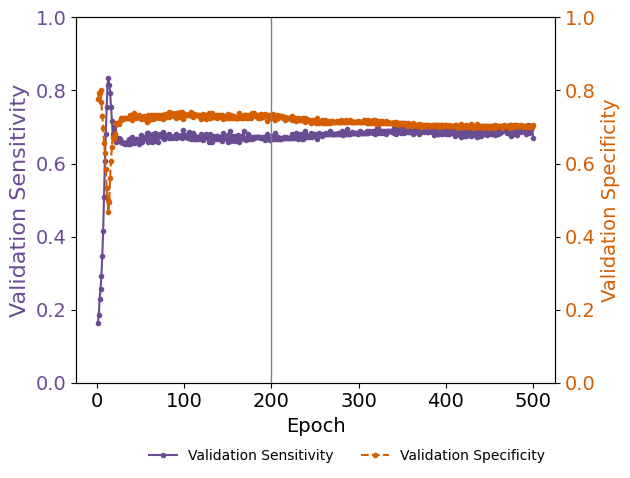}
    \end{subfigure}

    \caption{An example of a loss, accuracy, AUC, sensitivity, and specificity curve of the MIL models on a generic pre-processed dataset with 10 elements per bag. Plots shows the results from using the optimal parameters for the deep MIL model (top row), gated deep MIL model (second row), linear MIL model (third row), and extreme MIL model (bottom row). The vertical dotted lines indicate the 15-epoch point and the vertical straight lines indicate the 200-epoch point.}
    \label{fig:millosscurves}
\end{figure}

For the deep MIL model, we started with the generic pre-processed data and investigated combinations of different number of nodes --- various values below 256, 256, 512, 1024, and 2048 --- as well as different learning rates --- 0.0001, 0.001, 0.01, 0.1 --- while keeping a batch size of 10. Also, for every optimal number of nodes and learning rate combination, we investigated to see if keeping the number of nodes fixed and changing the learning rate to within the same order of magnitude would lead to any changes; we found no pertinent effects. We discovered that at below 256 nodes the models did not consistently perform well, that at 512 nodes the models performed a little bit better than at 256, and there was no noticeable difference between 512, 1024, and 2048 nodes. Hence, we chose 512 nodes with a learning rate of 0.001 for the deep MIL model. This translated to 362 nodes for the gated deep MIL model to ensure that the number of parameters were the same, for which we found the same optimal rate of 0.001. We then investigated the different batch sizes and found that at 50 and below there are instabilities in the loss curve, and as the batch size increases above 100, the rate of convergence decreases without any improvement in the validation AUC. Finally, we find that the loss curve and AUC on the validation set converges after 100 epochs. However, we find that stability of the sensitivity and specificity only occurs at around Epoch 200. Therefore, we stop the training at 200 epochs. For the linear MIL and extreme MIL models, we set the number of nodes to 512 and the number of epochs to 200 in order to keep the comparison fair but tuned the other parameters in a similar way as for the deep MIL model. This led to a learning rate of 0.0001 and a batch size of 100 for both. For the linear MIL model, larger learning rates led to higher instabilities in all the metrics, and learning rates of $10^{-5}$ did not always converge even after 500 epochs. An example of the training loop outcome for the four models can be seen in \autoref{fig:millosscurves}.

\begin{figure}[!ht]
    \centering
    \begin{subfigure}{0.49\textwidth}
        \centering
        \includegraphics[width=\linewidth]{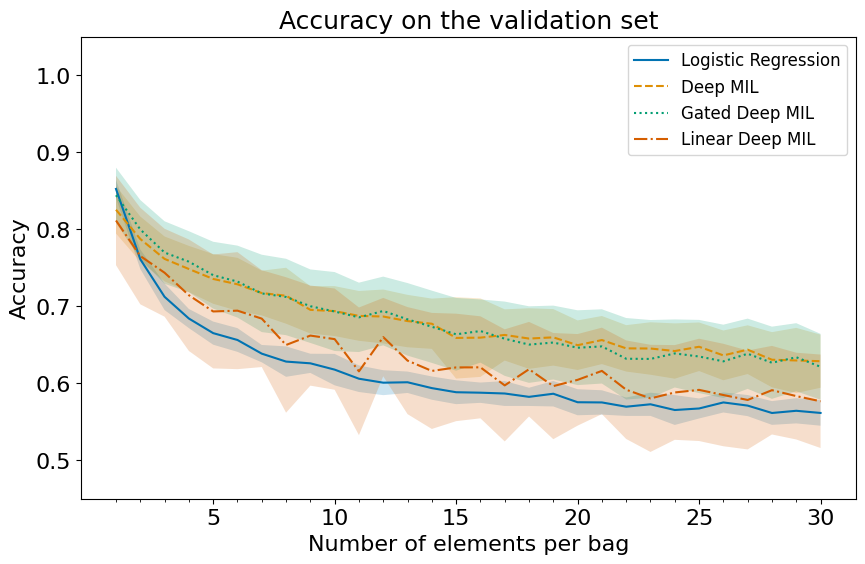}
    \end{subfigure}
    \hfill
    \begin{subfigure}{0.49\textwidth}
        \centering
        \includegraphics[width=\linewidth]{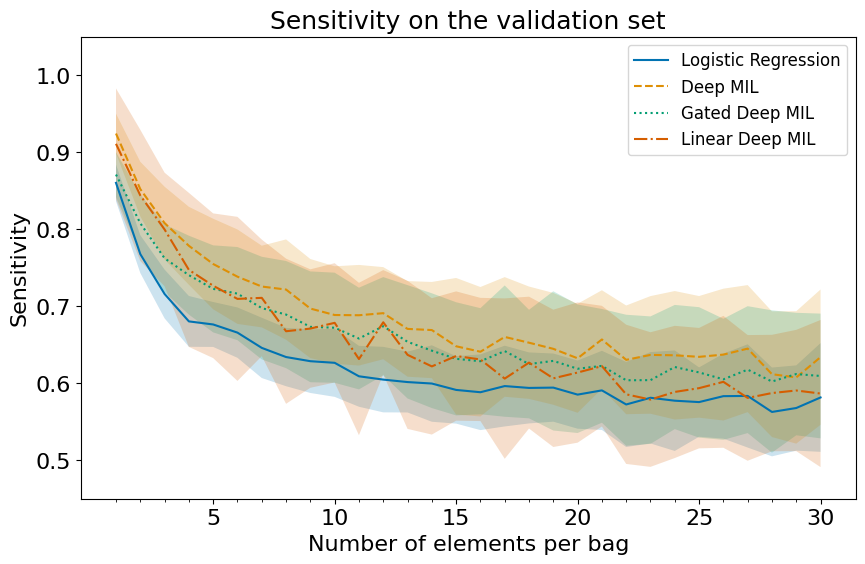}
    \end{subfigure}

    \vspace{0.1cm} 

    \begin{subfigure}{0.49\textwidth}
        \centering
        \includegraphics[width=\linewidth]{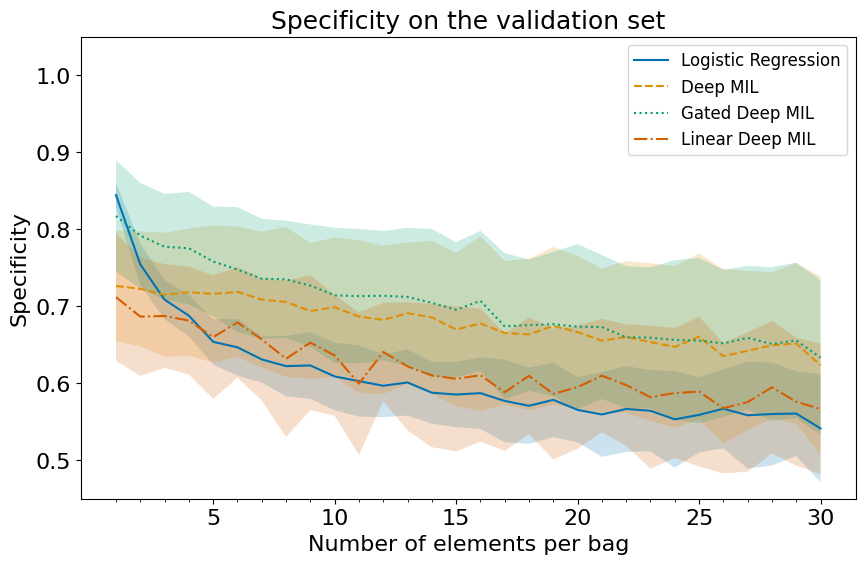}
    \end{subfigure}
    \hfill
    \begin{subfigure}{0.49\textwidth}
        \centering
        \includegraphics[width=\linewidth]{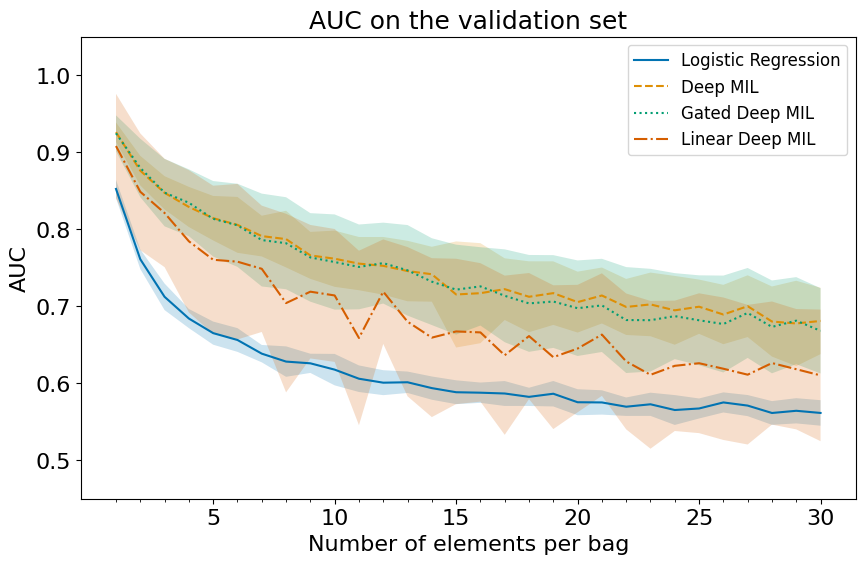}
    \end{subfigure}

    \caption{The accuracy, sensitivity, specificity, and AUC of the multiple instance learning models on the generic pre-processed data when stopping the deep MIL models after 15 epochs and the linear MIL model after 500 epochs. The results for logistic regression models are displayed with a solid blue line, for the deep MIL with a dashed orange line, for the gated deep MIL with a dotted green line, and for the linear MIL with a dash-dotted red line. The lines and bands for all models represent the calculated means and standard deviations respectively over the permutations of the bags over 20 different random permutations of the train-test split. We note that the significant widening of the standard deviations of the results from the deep and gated deep MIL models compared to those in \autoref{fig:deepmil} and \autoref{fig:linearmil} indicate that stopping early could lead to better models but at the expense of a considerable decrease in robustness.}
    \label{fig:earlystop}
\end{figure}

\autoref{fig:millosscurves} also shows the potential of obtaining slightly better results if stopping at an earlier epoch for the deep MIL models. We investigate this by re-running the algorithms but stopping the training after 15 epochs. In addition, we notice that for the linear MIL model, the results at Epoch 500 are potentially higher than that at 200. However, as shown in \autoref{fig:earlystop} we find that stopping the deep MIL models early only sometimes leads to better results while paying the price of being much noisier. We also find no major difference in the results of the linear MIL model.

\begin{figure}[!ht]
    \centering
    \begin{subfigure}{0.45\textwidth}
        \centering
        \includegraphics[width=\linewidth]{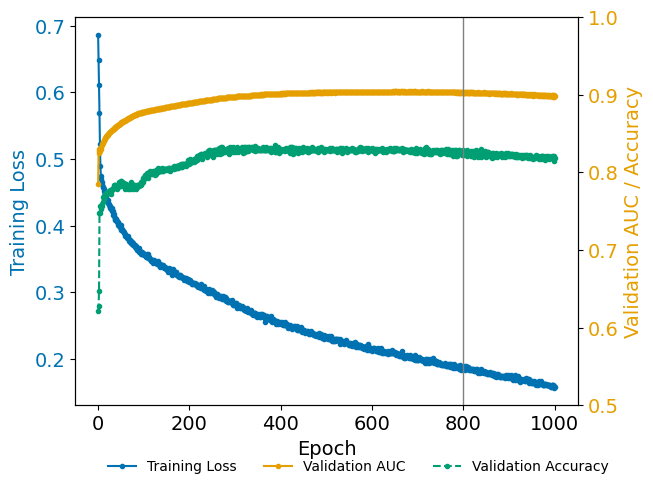}
    \end{subfigure}
    \hfill
    \begin{subfigure}{0.45\textwidth}
        \centering
        \includegraphics[width=\linewidth]{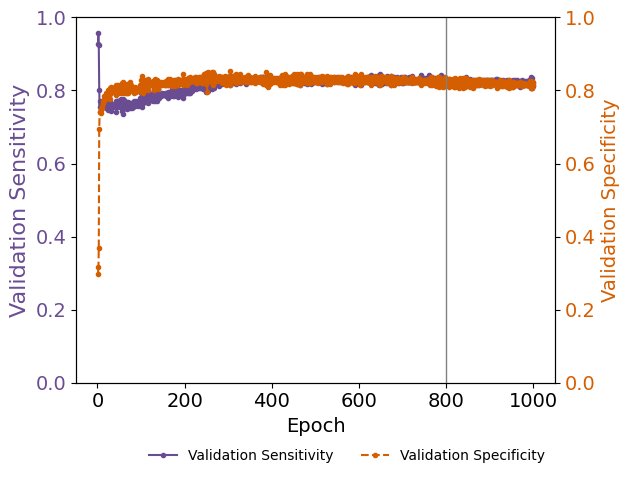}
    \end{subfigure}

    \vspace{0.1cm} 

    \begin{subfigure}{0.45\textwidth}
        \centering
        \includegraphics[width=\linewidth]{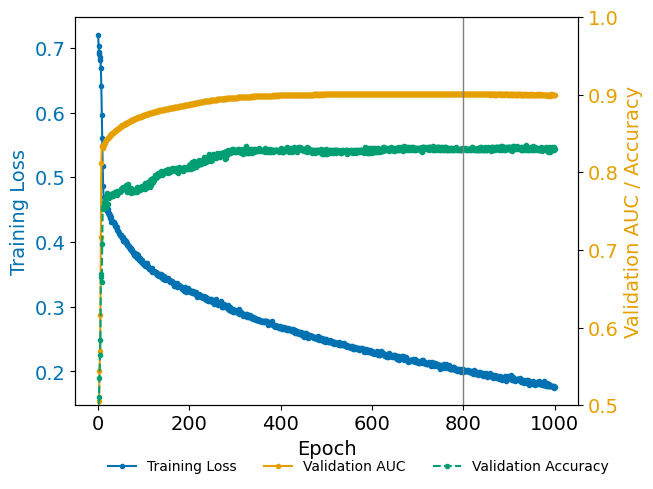}
    \end{subfigure}
    \hfill
    \begin{subfigure}{0.45\textwidth}
        \centering
        \includegraphics[width=\linewidth]{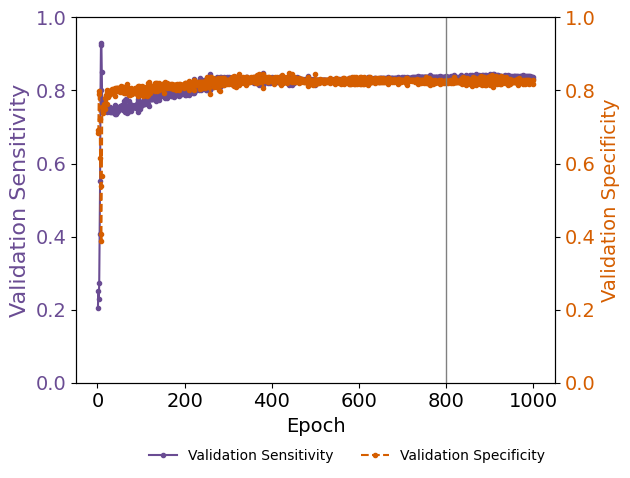}
    \end{subfigure}

    \begin{subfigure}{0.45\textwidth}
        \centering
        \includegraphics[width=\linewidth]{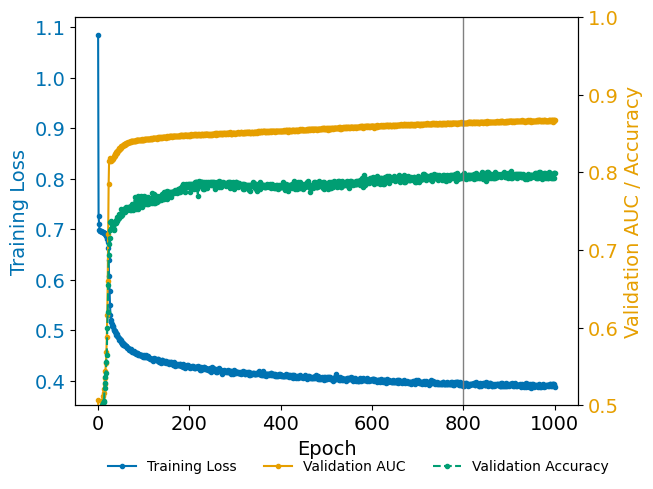}
    \end{subfigure}
    \hfill
    \begin{subfigure}{0.45\textwidth}
        \centering
        \includegraphics[width=\linewidth]{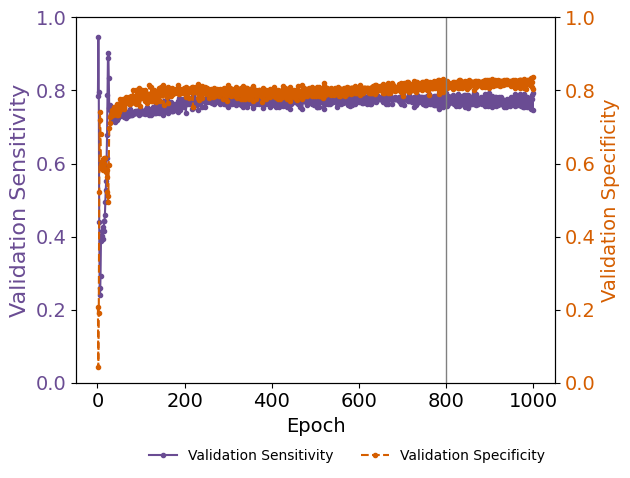}
    \end{subfigure}

    \begin{subfigure}{0.45\textwidth}
        \centering
        \includegraphics[width=\linewidth]{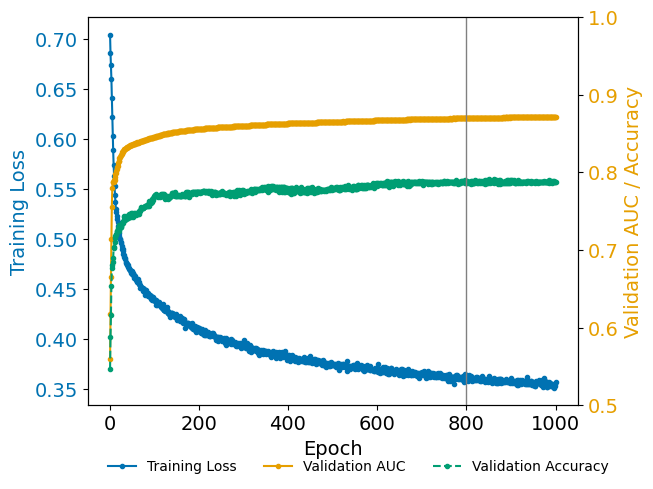}
    \end{subfigure}
    \hfill
    \begin{subfigure}{0.45\textwidth}
        \centering
        \includegraphics[width=\linewidth]{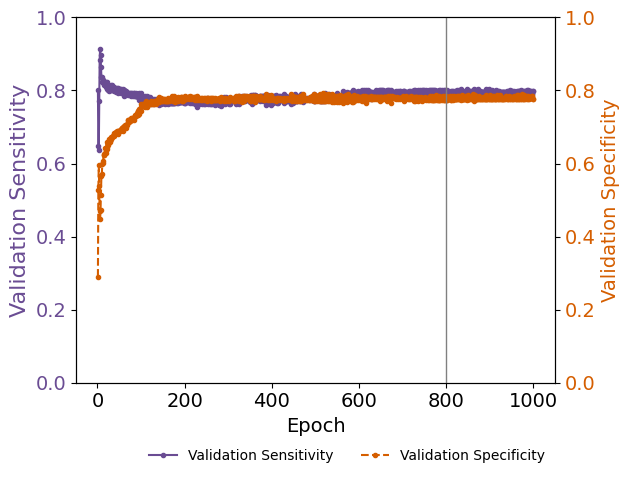}
    \end{subfigure}

    \caption{An example of a loss, accuracy, AUC, sensitivity, and specificity curve of the MIL models on a specialized pre-processed dataset with 10 elements per bag. Plots shows the results from using the optimal parameters for the deep MIL model (top row), gated deep MIL model (second row), linear MIL model (third row), and extreme MIL model (bottom row). The vertical straight lines indicate the 800-epoch point.}
    \label{fig:millosscurvesspecialized}
\end{figure}

To train the models for the specialized pre-processed data, we chose the optimal hyperparameter values for the generic pre-processed data as a starting point. We kept all factors the same apart from the learning rate and the number of epochs which we explored in a similar fashion to how it was done for the generic pre-processed data. This led to a learning rate of 0.0001 over 800 epochs for all models. The reason for this change in optimal hyperparameters between the types of datasets could be because when using the specialized pre-processing data, a lot of the learning is already done in the pre-processing step, and so the MIL models be fine-tuned more consistently over more epochs. Similar to \autoref{fig:millosscurves}, we present an example of the outcome of a training loop for the four models in \autoref{fig:millosscurvesspecialized} when using the specialized pre-processed data.

\section{The advantages of using the higher-dimension feature vector over the original feature vector in attention-based MIL models}
\label{sec:lowdimmil}
\begin{figure}[!ht]
    \centering
    \begin{subfigure}{0.49\textwidth}
        \centering
        \includegraphics[width=\linewidth]{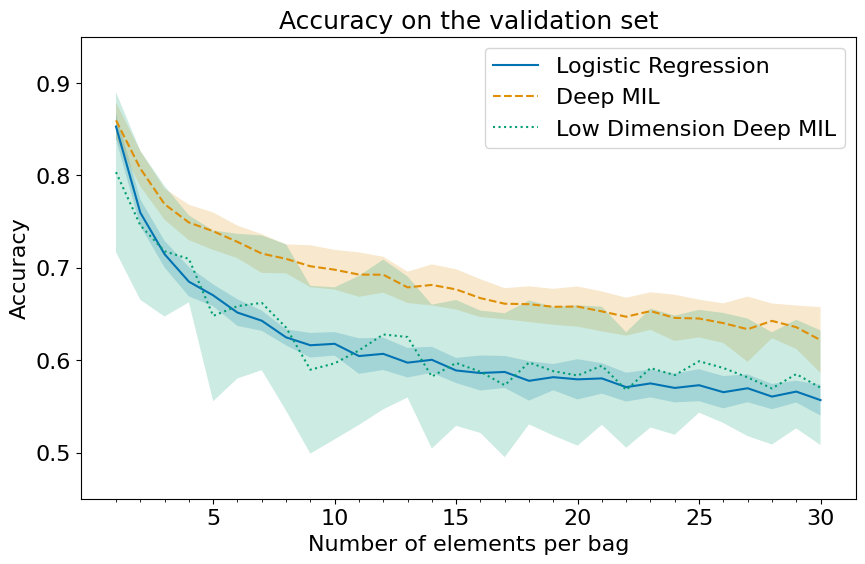}
    \end{subfigure}
    \hfill
    \begin{subfigure}{0.49\textwidth}
        \centering
        \includegraphics[width=\linewidth]{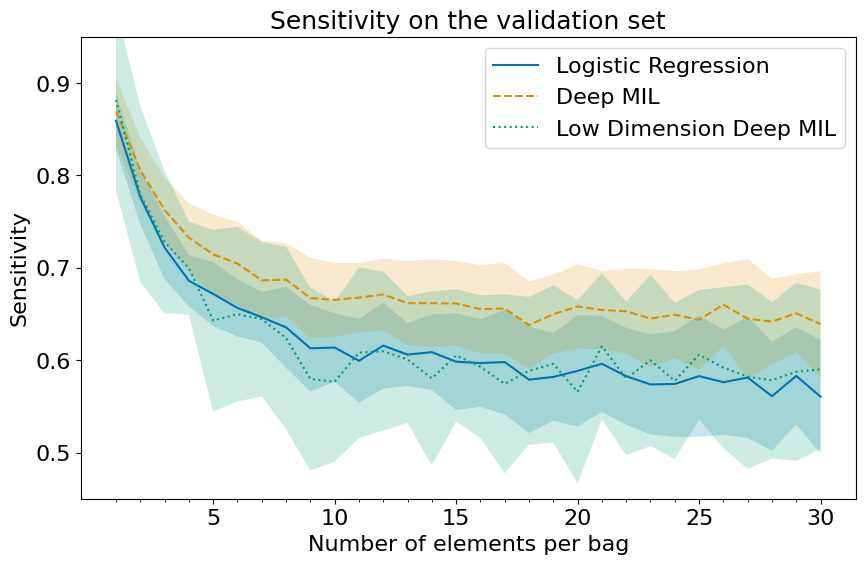}
    \end{subfigure}

    \vspace{0.1cm} 

    \begin{subfigure}{0.49\textwidth}
        \centering
        \includegraphics[width=\linewidth]{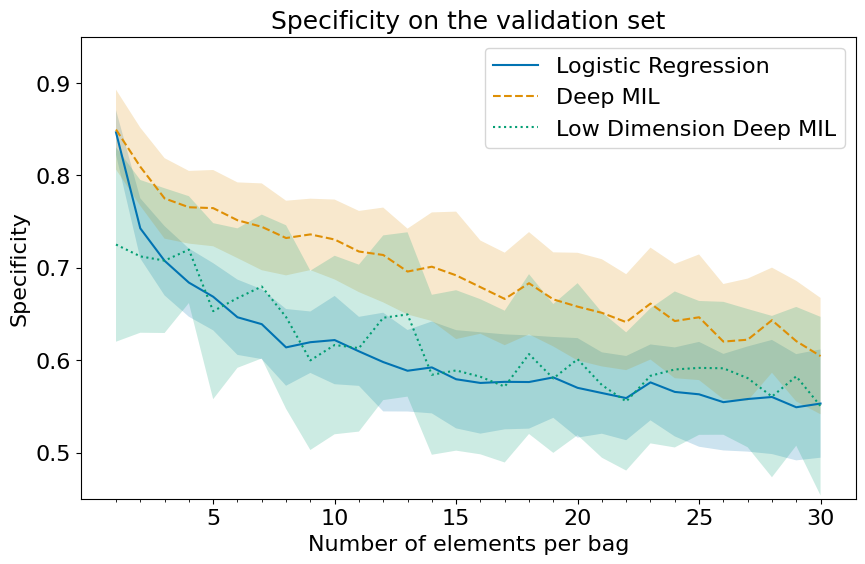}
    \end{subfigure}
    \hfill
    \begin{subfigure}{0.49\textwidth}
        \centering
        \includegraphics[width=\linewidth]{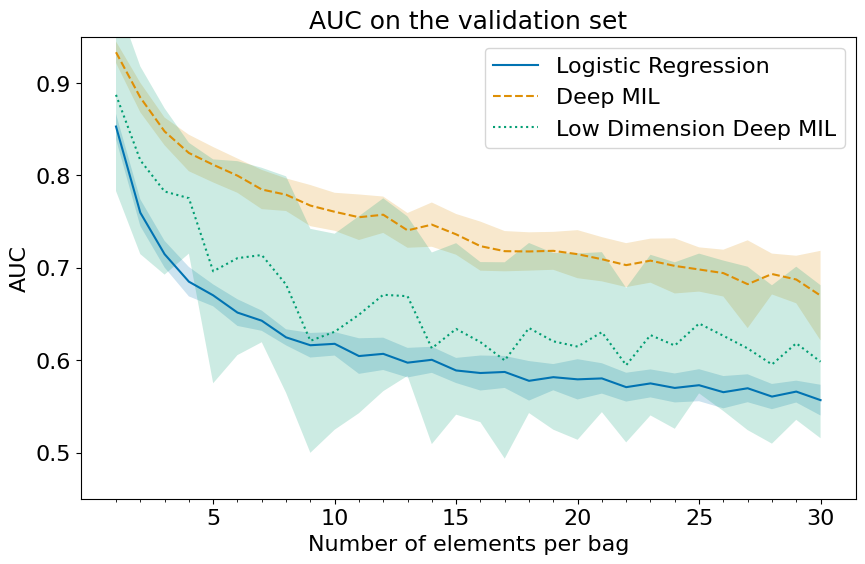}
    \end{subfigure}

    \caption{The accuracy, sensitivity, specificity, and AUC of the multiple instance learning models on the generic pre-processed data. The results for the logistic regression are displayed with a solid blue line, for the deep MIL with a dashed orange line, and for the low-dimensional deep MIL with a dotted green line. The lines and bands for all models represent the calculated means and standard deviations respectively over the permutations of the bags over 20 different random permutations of the train-validation split.}
    \label{fig:lowdim}
\end{figure}

In this section, we demonstrate the ineffectiveness of using the original feature vector in the aggregation step as introduced in \autoref{subsec:abdmil} and for which the results are displayed in  \autoref{fig:lowdim}. As can been seen in the plots, although sometimes the low dimensional deep MIL model performs well, it is highly volatile in its training and can sometimes perform poorly, including worse than the logistic regression baseline. This is the reason we have adopted the high-dimension feature vectors for our main results.

\section{Fine-tuning ResNET-18 to individual blood cell classification}
\label{sec:finetuneresnet}
\begin{figure}[!ht]
    \centering
    \includegraphics[width=\textwidth]{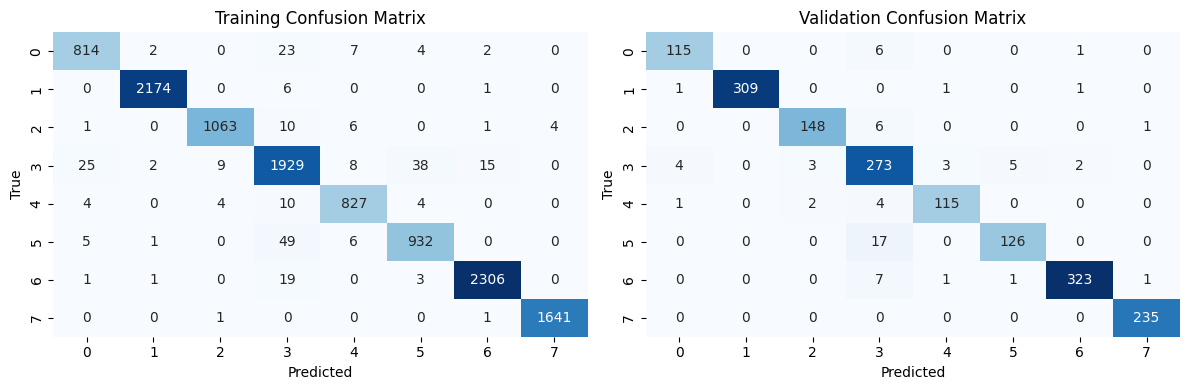}
    \caption{The confusion matrix on the training and test set after training the ResNET-18 architecture for 25 epochs on the training set of individual blood cells.}
    \label{fig:pretrainedconfmatrix}
\end{figure}

In this section, we present how the fine-tuned ResNET model that was used in the specialized pre-processing was trained. We used the 11,959 training images to train and the 1,712 validation images as the validation set. The goal of training the model was to create something that had >95\% accuracy on the validation set to mimic an ``ideal" pre-processing system for the eventual MIL problem.
We replaced the vanilla ResNET-18 network's last layer with 8 nodes (for the 8 different cell types) and used an Adam optimizer with a learning rate of 0.002 and a decay rate of 0.99. We found that after 25 epochs using a batch size of 30, we were able to obtain validation accuracies of >95\%. The confusion matrix generated using the final model can be seen in \autoref{fig:pretrainedconfmatrix}.

\section{Performance vs Number of Hidden Nodes}
\label{sec:hiddennodes}

\begin{figure}[h!]
    \centering
    \begin{subfigure}{0.49\textwidth}
        \centering
        \includegraphics[width=\linewidth]{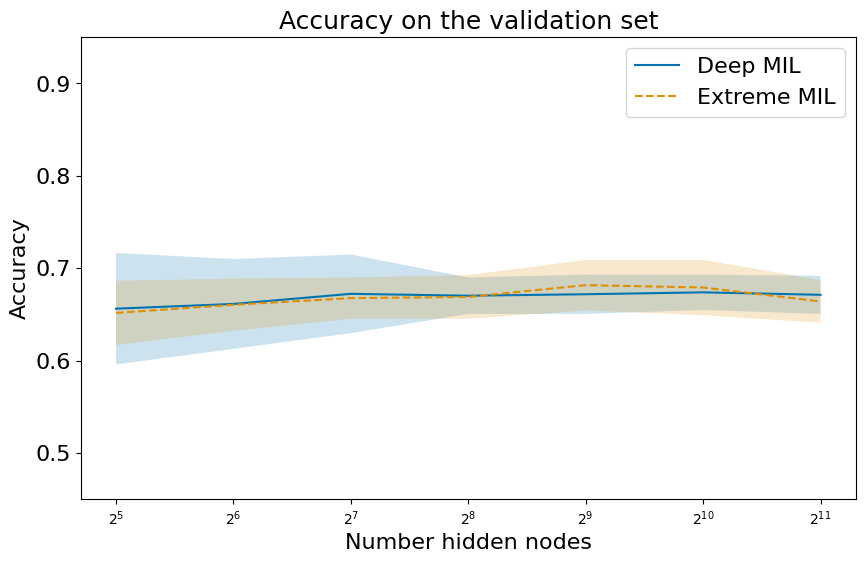}
    \end{subfigure}
    \hfill
    \begin{subfigure}{0.49\textwidth}
        \centering
        \includegraphics[width=\linewidth]{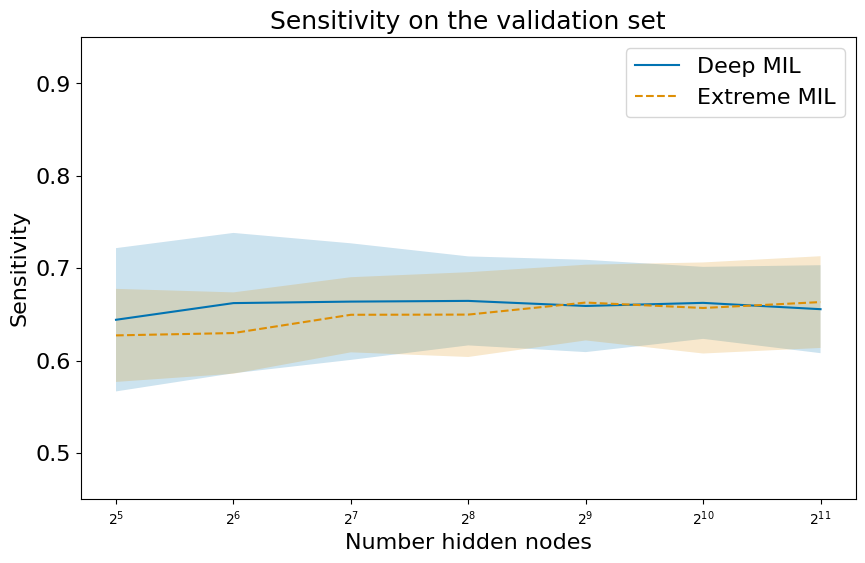}
    \end{subfigure}

    \vspace{0.1cm} 

    \begin{subfigure}{0.49\textwidth}
        \centering
        \includegraphics[width=\linewidth]{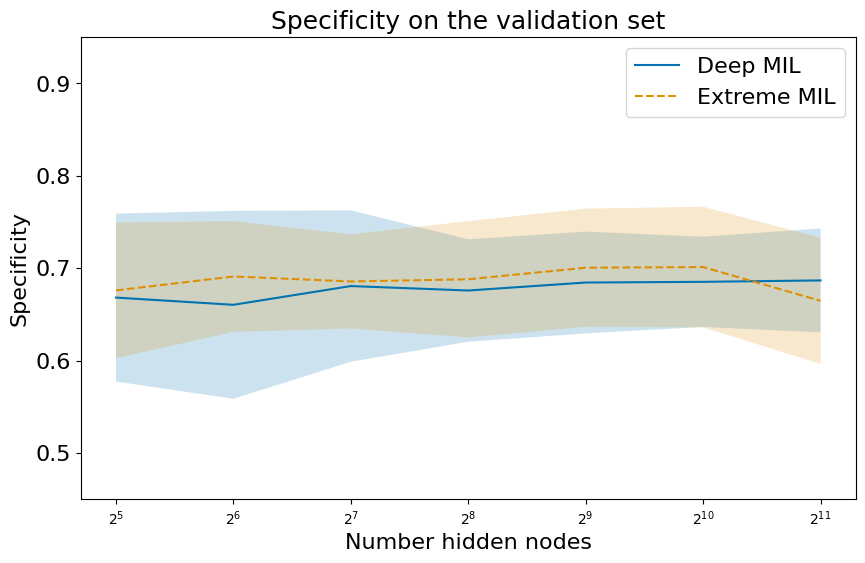}
    \end{subfigure}
    \hfill
    \begin{subfigure}{0.49\textwidth}
        \centering
        \includegraphics[width=\linewidth]{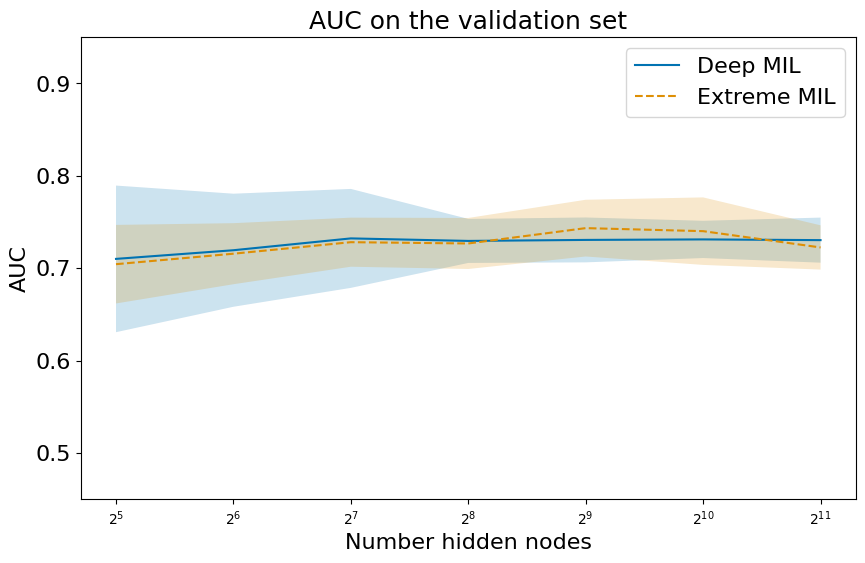}
    \end{subfigure}

    \caption{The effect that increasing the number of hidden nodes in the single layer of the attention model has on the performance of the deep and extreme MIL models with 15 elements per bag.}
    \label{fig:numhiddennodes}
\end{figure}

In this section, we demonstrate the results of investigating the effect that increasing the number of hidden nodes in the single layer of the attention model has on the performance of the deep and extreme MIL models. For this, we perform the experiment with a constant (15) number of generic pre-processed elements per bag, while varying the number of hidden nodes. The hyperparameters were tuned in a similar fashion to the method in \autoref{sec:hyperparameters}, with the batch size being 100 across all experiments. The final set of used parameters listed in \autoref{tab:numhiddennodes} and the results can be seen in \autoref{fig:numhiddennodes}. As mentioned in the main text in \autoref{sec:discussion}, we observed an improvement in model stability around $2^8$ nodes with the ELM architecture did not exhibit significantly worse performance at any hidden layer size, suggesting these performance gains do not have to be at the expense of the number of trained parameters.

\begin{table}[!ht]
    \centering
    \begin{tabular}{|l|c|c|c|}
        \toprule
        Model & Number of hidden nodes & Learning rate & Number of epochs \\
        \midrule
        Deep MIL & 32 & 0.0005 & 200 \\
        Extreme MIL & 32 & 0.001 & 300 \\
        Deep MIL & 64 & 0.0001 & 400 \\
        Extreme MIL & 64 & 0.001 & 300 \\
        Deep MIL & 128 & 0.0001 & 400 \\
        Extreme MIL & 128 & 0.001 & 300 \\
        Deep MIL & 256 & 0.001 & 200 \\
        Extreme MIL & 256 & 0.0005 & 500 \\
        Deep MIL & 512 & 0.001 & 200 \\
        Extreme MIL & 512 & 0.0001 & 500 \\
        Deep MIL & 1024 & 0.001 & 200 \\
        Extreme MIL & 1024 & 0.0001 & 200 \\
        Deep MIL & 2048 & 0.001 & 200 \\
        Extreme MIL & 2048 & 0.001 & 200 \\
        \bottomrule
    \end{tabular}
    \caption{The selection of hyperparameters when investigating the effect of the number of hidden nodes in the deep and extreme MIL models. All models used a bath size of 100.}
    \label{tab:numhiddennodes}
\end{table}

\begin{table}[!ht]
\centering
\begin{tabular}{@{}llllllll@{}}
\toprule
Model 1             & Model 2             & Metric      & Delta (\%) & St. dev (\%) & p-value            &   &  \\ \midrule
Gated deep MIL      & Deep MIL            & Accuracy    & 0.44       & 0.73         & 1.000              &   &  \\
(specialized)       & (specialized)       & Sensitivity & 0.16       & 0.64         & 1.000              &   &  \\
                    &                     & Specificity & 0.71       & 0.91         & 1.000              &   &  \\
                    &                     & AUC         & 0.46       & 0.91         & 1.000              &   &  \\
Gated deep MIL      & Gated deep MIL      & Accuracy    & 13.01      & 2.25         & p \textless 0.0017 & * &  \\
(specialized)       & (generic)           & Sensitivity & 11.45      & 1.35         & p \textless 0.0017 & * &  \\
                    &                     & Specificity & 14.57      & 4.27         & p \textless 0.0017 & * &  \\
                    &                     & AUC         & 13.94      & 3.21         & p \textless 0.0017 & * &  \\
Gated deep MIL      & Logistic regression & Accuracy    & 16.55      & 4.15         & p \textless 0.0017 & * &  \\
(specialized)       & (specialized)       & Sensitivity & 14.19      & 3.49         & p \textless 0.0017 & * &  \\
                    &                     & Specificity & 18.91      & 4.94         & p \textless 0.0017 & * &  \\
                    &                     & AUC         & 23.70      & 4.43         & p \textless 0.0017 & * &  \\
Deep MIL            & Deep MIL            & Accuracy    & 13.36      & 2.08         & p \textless 0.0017 & * &  \\
(specialized)       & (generic)           & Sensitivity & 11.66      & 1.93         & p \textless 0.0017 & * &  \\
                    &                     & Specificity & 15.06      & 3.40         & p \textless 0.0017 & * &  \\
                    &                     & AUC         & 14.34      & 2.99         & p \textless 0.0017 & * &  \\
Deep MIL            & Logistic regression & Accuracy    & 16.11      & 3.97         & p \textless 0.0017 & * &  \\
(specialized)       & (specialized)       & Sensitivity & 14.02      & 3.30         & p \textless 0.0017 & * &  \\
                    &                     & Specificity & 18.20      & 4.77         & p \textless 0.0017 & * &  \\
                    &                     & AUC         & 23.24      & 4.24         & p \textless 0.0017 & * &  \\
Gated deep MIL      & Deep MIL            & Accuracy    & 0.79       & 0.50         & 0.005              &   &  \\
(generic)           & (generic)           & Sensitivity & 0.38       & 0.98         & 0.921              &   &  \\
                    &                     & Specificity & 1.20       & 1.15         & 0.037              &   &  \\
                    &                     & AUC         & 0.86       & 0.52         & 0.016              &   &  \\
Gated deep MIL      & Logistic regression & Accuracy    & 7.87       & 1.64         & p \textless 0.0017 & * &  \\
(generic)           & (generic)           & Sensitivity & 6.00       & 2.10         & p \textless 0.0017 & * &  \\
                    &                     & Specificity & 9.75       & 2.05         & p \textless 0.0017 & * &  \\
                    &                     & AUC         & 14.09      & 1.65         & p \textless 0.0017 & * &  \\
Deep MIL            & Logistic regression & Accuracy    & 7.09       & 1.59         & p \textless 0.0017 & * &  \\
(generic)           & (generic)           & Sensitivity & 5.62       & 1.53         & p \textless 0.0017 & * &  \\
                    &                     & Specificity & 8.55       & 2.50         & p \textless 0.0017 & * &  \\
                    &                     & AUC         & 13.23      & 1.54         & p \textless 0.0017 & * &  \\
Logistic regression & Logistic regression & Accuracy    & 4.33       & 1.11         & p \textless 0.0017 & * &  \\
(specialized)       & (generic)           & Sensitivity & 3.27       & 1.57         & p \textless 0.0017 & * &  \\
                    &                     & Specificity & 5.40       & 1.22         & p \textless 0.0017 & * &  \\
                    &                     & AUC         & 4.33       & 1.11         & p \textless 0.0017 & * &  \\ \bottomrule
\end{tabular}
    \caption{Statistical comparison of model performance curves illustrated in \autoref{fig:deepmil} comparing deep MIL with gated MIL. Delta (\%) indicates the percentage improvement in model accuracy, sensitivity, specificity, and AUC when choosing Model 1 over Model 2, averaged over all bag sizes. A p-cutoff of 0.05 was corrected using the Bonferroni method, resulting in significance below p = 0.0017.}
    \label{tab:fig6_stats}
\end{table}

\begin{table}[!ht]
\centering
\begin{tabular}{@{}lllllll@{}}
\toprule
Model 1             & Model 2             & Metric      & Delta (\%) & St. dev (\%) & p-value            &   \\ \midrule
Deep MIL            & Linear MIL          & Accuracy    & 4.31       & 1.99         & p \textless 0.0017 & * \\
(specialized)       & (specialized)       & Sensitivity & 3.73       & 2.22         & p \textless 0.0017 & * \\
                    &                     & Specificity & 4.89       & 1.92         & p \textless 0.0017 & * \\
                    &                     & AUC         & 4.33       & 2.45         & p \textless 0.0017 & * \\
Deep MIL            & Deep MIL            & Accuracy    & 13.36      & 2.08         & p \textless 0.0017 & * \\
(specialized)       & (generic)           & Sensitivity & 11.66      & 1.93         & p \textless 0.0017 & * \\
                    &                     & Specificity & 15.06      & 3.4          & p \textless 0.0017 & * \\
                    &                     & AUC         & 14.34      & 2.99         & p \textless 0.0017 & * \\
Deep MIL            & Logistic regression & Accuracy    & 16.11      & 3.97         & p \textless 0.0017 & * \\
(specialized)       & (specialized)       & Sensitivity & 14.02      & 3.3          & p \textless 0.0017 & * \\
                    &                     & Specificity & 18.2       & 4.77         & p \textless 0.0017 & * \\
                    &                     & AUC         & 23.24      & 4.24         & p \textless 0.0017 & * \\
Linear MIL          & Linear MIL          & Accuracy    & 12.36      & 2.45         & p \textless 0.0017 & * \\
(specialized)       & (generic)           & Sensitivity & 9.69       & 2.55         & p \textless 0.0017 & * \\
                    &                     & Specificity & 15.04      & 2.61         & p \textless 0.0017 & * \\
                    &                     & AUC         & 14.23      & 3.69         & p \textless 0.0017 & * \\
Linear MIL          & Logistic regression & Accuracy    & 11.8       & 2.98         & p \textless 0.0017 & * \\
(specialized)       & (specialized)       & Sensitivity & 10.29      & 2.49         & p \textless 0.0017 & * \\
                    &                     & Specificity & 13.31      & 3.95         & p \textless 0.0017 & * \\
                    &                     & AUC         & 18.9       & 3.24         & p \textless 0.0017 & * \\
Deep MIL            & Linear MIL          & Accuracy    & 3.31       & 1.83         & p \textless 0.0017 & * \\
(generic)           & (generic)           & Sensitivity & 1.75       & 3.45         & p \textless 0.0017 & * \\
                    &                     & Specificity & 4.87       & 2.02         & p \textless 0.0017 & * \\
                    &                     & AUC         & 4.22       & 2.48         & p \textless 0.0017 & * \\
Deep MIL            & Logistic regression & Accuracy    & 7.09       & 1.59         & p \textless 0.0017 & * \\
(generic)           & (generic)           & Sensitivity & 5.62       & 1.53         & p \textless 0.0017 & * \\
                    &                     & Specificity & 8.55       & 2.5          & p \textless 0.0017 & * \\
                    &                     & AUC         & 13.23      & 1.54         & p \textless 0.0017 & * \\
Linear MIL          & Logistic regression & Accuracy    & 3.77       & 1.96         & p \textless 0.0017 & * \\
(generic)           & (generic)           & Sensitivity & 3.87       & 2.63         & p \textless 0.0017 & * \\
                    &                     & Specificity & 3.67       & 2.53         & p \textless 0.0017 & * \\
                    &                     & AUC         & 9.01       & 2.88         & p \textless 0.0017 & * \\
Logistic regression & Logistic regression & Accuracy    & 4.33       & 1.11         & p \textless 0.0017 & * \\
(specialized)       & (generic)           & Sensitivity & 3.27       & 1.57         & p \textless 0.0017 & * \\
                    &                     & Specificity & 5.4        & 1.22         & p \textless 0.0017 & * \\
                    &                     & AUC         & 4.33       & 1.11         & p \textless 0.0017 & * \\ \bottomrule
\end{tabular}
    \caption{Statistical comparison of model performance curves from \autoref{fig:linearmil} comparing deep MIL with linear MIL. Delta (\%) indicates the percentage improvement in model accuracy, sensitivity, specificity, and AUC when choosing Model 1 over Model 2, averaged over all bag sizes. A p-cutoff of 0.05 was corrected using the Bonferroni method, resulting in significance below p = 0.0017.}
    \label{tab:fig7_stats}
\end{table}

\begin{table}[!ht]
\centering
\begin{tabular}{@{}lllllll@{}}
\toprule
Model 1       & Model 2       & Metric      & Delta (\%) & St. dev (\%) & p-value            &   \\ \midrule
Deep MIL      & Extreme MIL   & Accuracy    & 1.81       & 0.73         & p \textless 0.0017 & * \\
(specialized) & (specialized) & Sensitivity & 2.02       & 1.03         & p \textless 0.0017 & * \\
              &               & Specificity & 1.61       & 0.8          & 0.4701             &   \\
              &               & AUC         & 1.47       & 0.97         & p \textless 0.0017 & * \\
Deep MIL      & Deep MIL      & Accuracy    & 13.36      & 2.08         & p \textless 0.0017 & * \\
(specialized) & (generic)     & Sensitivity & 11.66      & 1.93         & p \textless 0.0017 & * \\
              &               & Specificity & 15.06      & 3.4          & p \textless 0.0017 & * \\
              &               & AUC         & 14.34      & 2.99         & p \textless 0.0017 & * \\
Extreme MIL   & Extreme MIL   & Accuracy    & 11.87      & 1.12         & p \textless 0.0017 & * \\
(specialized) & (generic)     & Sensitivity & 9.57       & 2.62         & p \textless 0.0017 & * \\
              &               & Specificity & 14.18      & 1.44         & p \textless 0.0017 & * \\
              &               & AUC         & 13.07      & 2.33         & p \textless 0.0017 & * \\
Deep MIL      & Extreme MIL   & Accuracy    & 0.32       & 1.09         & 0.2612             &   \\
(generic)     & (generic)     & Sensitivity & -0.08      & 2.68         & p \textless 0.0017 & * \\
              &               & Specificity & 0.73       & 4.24         & p \textless 0.0017 & * \\
              &               & AUC         & 0.2        & 0.7          & 0.9981             &   \\ \bottomrule
\end{tabular}
    \caption{Statistical comparison of model performance curves from \autoref{fig:emil} comparing deep MIL with extreme MIL. Delta (\%) indicates the percentage improvement in model accuracy, sensitivity, specificity, and AUC when choosing Model 1 over Model 2, averaged over all bag sizes. A p-cutoff of 0.05 was corrected using the Bonferroni method, resulting in significance below p = 0.0017.}
    \label{tab:fig8_stats}
\end{table}

\end{document}